%% file: clonescaper.tex
\documentclass[10pt,journal,compsoc]{IEEEtran}

\ifCLASSOPTIONcompsoc
  \usepackage[nocompress]{cite}
\else
  \usepackage{cite}
\fi

\ifCLASSINFOpdf

\else
  
\fi

\usepackage{booktabs}
\usepackage{diagbox}
\usepackage{subcaption} 
\usepackage[linesnumbered,ruled]{algorithm2e}
\usepackage{xspace}
\makeatletter
\renewcommand{\@algocf@capt@plain}{above}
\makeatother
\usepackage{algorithmicx}
\usepackage{algpseudocode}    
\usepackage{amsmath}
    
\usepackage{amssymb}
\usepackage{bbding} 
\usepackage{calc}
\usepackage{markdown}
\usepackage{framed} 
\usepackage{tcolorbox} 
\usepackage{circuitikz}
\usepackage{color}            
\usepackage{etex}
\usepackage{etoolbox}
\usepackage[justification=centering]{caption}
\usepackage{enumitem}
\usepackage{filecontents}
\usepackage{graphicx}
\usepackage[customcolors]{hf-tikz}
\usepackage{listings}
\usepackage{microtype}        
\usepackage{multirow}
\usepackage{pgfplots}
\usepackage{pifont}
\usepackage{pst-circ}
\usepackage{pstricks}
\usepackage{relsize}
\usepackage{amsmath}
\usepackage{setspace}
\usepackage{stmaryrd}
\usepackage[T1]{fontenc}      
\usepackage{tikz}
\usepackage[utf8]{inputenc}   
\usepackage[USenglish]{babel} 
\usetikzlibrary{matrix,arrows,positioning,shapes,backgrounds,fit}
\pgfplotsset{compat=1.3}
\usepackage{wasysym}
\usepackage{xcolor}
\usepackage{ulem}
\usepackage{url}
\usepackage{balance}
\newcommand\ignore[1]{}

\newcommand{\ourtool}{\textsc{CloneGen}\xspace}
\newcommand{\DRLSG}{\textsc{DRLSG}\xspace}

\newcommand{\NICARD}{\textsc{NiCad}\xspace}
\newcommand{\CCAligner}{\textsc{CCAligner}\xspace}
\newcommand{\CCLearner}{\textsc{CCLearner}\xspace}
\newcommand{\Deckard}{\textsc{Deckard}\xspace}

\newcommand{\Astnn}{\textsc{ASTNN}\xspace}
\newcommand{\FCDector}{\textsc{FCDetector}\xspace}
\newcommand{\SourcererCC}{\textsc{SourcererCC}\xspace}
\newcommand{\TBCCD}{\textsc{TBCCD}\xspace}
\newcommand{\CCGraph}{\textsc{CCGraph}\xspace}
\newcommand{\CCFinderX}{\textsc{CCFinderX}\xspace}
\newcommand{\TokenLSTM}{\textsc{TextLSTM}\xspace}
\newcommand{\TextLSTM}{\textsc{TextLSTM}\xspace}

\let\oldnl\nl
\newcommand{\nonl}{\renewcommand{\nl}{\let\nl\oldnl}}%
\newcommand*{\priority}[1]{\begin{tikzpicture}[scale=0.11]%
		\draw (0,0) circle (1);
		\fill[fill opacity=0.5,fill=black] (0,0) -- (180:1) arc (180:180-#1*3.6:1) -- cycle;
\end{tikzpicture}}

\newtheorem{myDef}{Definition}

\definecolor{darkblue}{rgb}{0, 0.125, 0.576}
\definecolor{dkgreen}{rgb}{0,0.6,0}
\definecolor{gray}{rgb}{0.5,0.5,0.5}
\definecolor{mauve}{rgb}{0.58,0,0.82}

\newcommand\xyx[1]{\textcolor{black}{{#1}}}

\newcommand{\codeff}[1]{\texttt{\small #1}}
\newcommand{\tool}[1]{\textsc{Clonescaper}}

\newcommand\rev[1]{\textcolor{black}{{#1}}}

\begin{document}
\normalem

\title{\rev{Challenging} Machine Learning-based  Clone  Detectors via  Semantic-preserving Code Transformations}

\author{Weiwei~Zhang,
        Shengjian~Guo,
        Hongyu~Zhang,
        Yulei~Sui,
        Yinxing~Xue,
        and Yun~Xu
\IEEEcompsocitemizethanks{
\IEEEcompsocthanksitem W.~Zhang is a master student with school of Computer Science and Technology, University of Science and Technology of China. wwzh@mail.ustc.edu.cn
\IEEEcompsocthanksitem Y.~Xue and Y.~Xu work in the School of Computer Science and Technology at the University of Science and Technology of China. (yxxue,xuyun)@ustc.edu.cn 
\IEEEcompsocthanksitem S.~G works in the Baidu Security. sjguo@baidu.com 
\IEEEcompsocthanksitem H.~Zhang works in the University of Newcastle. Hongyu.Zhang@newcastle.edu.au 
\IEEEcompsocthanksitem Y.~Sui works in the University of Technology Sydney. yulei.sui@uts.edu.au\protect
}

\thanks{(Corresponding author: Yinxing~Xue.)\\Manuscript received April 19, 2005; revised August 26, 2015.}}

\markboth{IEEE TRANSACTIONS ON SOFTWARE ENGINEERING}%
{Shell \MakeLowercase{\textit{et al.}}: Bare Demo of IEEEtran.cls for Computer Society Journals}

\IEEEtitleabstractindextext{%
\input{abstract}

}

\maketitle

\IEEEdisplaynontitleabstractindextext

\IEEEpeerreviewmaketitle

\input{introduction}

\input{motivation}

\input{approach}

\input{operator}

\input{deepRL}

\input{evaluation}

\input{related}

\input{conclusion}






\balance 
\ifCLASSOPTIONcaptionsoff
  \newpage
\fi

\bibliographystyle{IEEEtran}
\bibliography{clonescaper}






\end{document}

%% file: abstract.tex
\begin{abstract}
Software clone detection identifies similar or identical code snippets. It has been an active research topic that attracts extensive attention over the last two decades. In recent years, machine learning (ML) based detectors, especially deep learning-based ones, have demonstrated impressive capability on clone detection. It seems that this longstanding problem has already been tamed owing to the advances in ML techniques. In this work, we would like to challenge the robustness of the recent ML-based clone detectors through code semantic-preserving transformations.
We first utilize fifteen simple code transformation operators combined with commonly-used heuristics \rev{(i.e., Random Search, Genetic Algorithm, and Markov Chain Monte Carlo)} to perform equivalent program transformation.  
\rev{Furthermore, we propose a deep reinforcement learning-based sequence generation (DRLSG) strategy to effectively guide the search process of generating clones that could escape from the detection.}
We then evaluate the ML-based detectors with the pairs of original and generated clones. 
We realize our method in a framework named \ourtool (stands for Clone Generator).   
\rev{In evaluation, we challenge the \rev{two} state-of-the-art ML-based detectors and four traditional detectors with the code clones after semantic-preserving transformations via the aid of \ourtool. }
Surprisingly, our experiments show that, despite the notable successes achieved by existing clone detectors, the ML models inside these detectors still cannot distinguish numerous clones produced by the code transformations in \ourtool. In addition, adversarial training of ML-based clone detectors using clones generated by \ourtool can improve their robustness and accuracy. \rev{Meanwhile, compared with  the commonly-used heuristics, the DRLSG strategy has shown the best effectiveness in generating code clones to decrease the detection accuracy of the ML-based detectors.}
Our investigation reveals an explicable but always ignored robustness issue of the latest ML-based detectors. Therefore, we call for more attention to the robustness of these new ML-based detectors. 
\end{abstract}

%% file: introduction.tex
{\section{Introduction}\label{sec:intro}}
\IEEEPARstart{C}{ode} reuse via copy-and-paste actions is common in software development. Such 
practice typically generates a large amount of similar code, which is often 
called \emph{code clones}. According to the large-scale study conducted by 
Mockus~\cite{Mockus2007}, \rev{more than 50\% of files were reused in some open-source projects.}
Though code cloning may be helpful under proper utilization~\cite{Kapser2008}, 
it can also become bad programming practice because of the painful maintenance 
costs~\cite{Bruno1997}. For example, Li et al.~\cite{Zhenmin2006} reported that 
22.3\% of operating systems' defects were introduced by code cloning. 
Moreover, code cloning also brings intelligence property violations
\cite{moss,Liu2006,Wang2009} and security problems~\cite{Zhenmin2006,Islam2016}.

In retrospect, source code clone detection is an active research domain that 
attracts extensive attentions. Multiple clone detectors have been proposed 
based on various types of code representations, including \emph{Textual}- or \emph{Token}-based~\cite{seunghak2005,WangSWXR18}, \emph{Ast}-based~\cite{JiangMSG07}, \emph{PDG}-based~\cite{Zou2021}, etc. 
These traditional detectors are primarily designed for syntax-based clone detection 
(clones with syntactic similarity). Besides, many of them are of limited capability
(e.g., specialized for a certain type of clones) and low-efficiency
\cite{roy2009comparison,walker2020open}.

Recently, the latest machine learning (ML) methods have been significantly enhancing the capabilities of clone detectors.  For example, \FCDector~\cite{FangLS0S20} trains 
a DNN network to detect clones of functions by capturing code syntax and semantic 
information through \emph{AST} (abstract syntax tree) and \emph{CFG} (control flow 
graph). \Astnn~\cite{Zhang2019} maintains an \emph{AST}-based neural source code representation that utilizes a bidirectional model to exploit the naturalness of source code statements for clone detection. \TBCCD~\cite{YuLCLXW19} obtains 
structural information of code fragments from \emph{AST} and lexical information 
from code tokens and adopts a tree-based traditional approach to detect semantic 
clones. \CCLearner~\cite{Liuqing2017} extracts tokens from source code to train 
a DNN-based classifier and then uses the classifier to detect clones in a  
codebase. These new ML-based approaches fuse the latest ML techniques with the code features extracted from clones, thus achieving highly accurate results. 
Usually, they can detect most semantic clones with accuracy over 
95\%~\cite{FangLS0S20,Zhang2019,YuLCLXW19}.

Though the new ML-based clone detectors have shown impressive accomplishments, 
their effectiveness heavily relies on well-labeled training data~\cite{zhu2016we}. 
The performance of a detector trained with one dataset can be less effective in detecting code clones in another dataset. 
Unlike simple texts, source code contains both textual and structural information, 
which makes ground-truth clone pairs more versatile. Hence, building a robust prediction model for code clone detection is inherently challenging.
For example, 
code fragments \codeff{int b=0;} and \codeff{int b;b=0;} have the same semantics, but their \emph{CFG} and \emph{AST} slightly differ. 
Hence, rather than applying heavy code obfuscations and compiler optimizations, can 
we merely \rev{generate} \rev{light-weight} source code variations at some program 
locations to \rev{effectively} lower the accuracy of a ML-based detector? 
From this point, we investigate and observe that \emph{adopting semantic-preserving 
transformation of a code fragment is of practical importance in validating the robustness of ML-based clone detectors.} 

Specifically, we \xyx{present a framework, named \ourtool, that 
performs semantic-preserving  
code transformations to challenge} ML-based clone detectors.
In \ourtool, we have developed 15 lightweight and semantic-preserving  code transformations 
(e.g., variable renaming, \codeff{for-loop} to \codeff{while-loop} conversion, 
code order swapping, etc.).
In general, \ourtool targets the \rev{cheap} yet effective transformation (or a 
combination of transformations)   on a given code snippet to evade the clone 
detection. 
 \rev{To effectively guide the combinations of the 15 atomic transformations, \ourtool supports various heuristic strategies (i.e., Random Search, Genetic Algorithm, and Markov Chain Monte Carlo). Essentially, to quickly find the (near-) optimal solutions for evading clone detection is an optimization problem of how to combine multiple  transformations. \rev{To address this, we design 
a deep reinforcement learning (DRL) sequence generation  model (called DRLSG), which uses a 
Proximal Policy Optimization (PPO) strategy neural network~\cite{schulman2017proximal}
to guide the search process}}.

\xyx{With \ourtool}, we select unique code snippets from the widely used OJClone
datasets \xyx{~\cite{MouLZWJ16}}. Given a selected code snippet $x$, we generate semantic-preserved variants and pair $x$ with each of them. Then, we feed the formed clone pairs to the ML-based detectors to \rev{challenge} whether they can still determine these code clones produced by our framework.

\rev{To evaluate the effectiveness of \ourtool, clone pairs generated by \ourtool are provided to the two state-of-the-art open-source ML-based detectors (namely, \Astnn and \TBCCD), as well as a baseline detector \TokenLSTM~\cite{xingjian2015convolutional}.} 
\rev{We  find that  40.1\%, 32.9\% and 44.2\% of the clone pairs supplied by \ourtool can successfully \xyx{evade} the detection of \Astnn, \TBCCD, and \TokenLSTM, respectively. Considering the original accuracy of these detectors (more than 95\%) on the OJClone datasets,  experimental results prove that \ourtool is notably effective in lowering the accuracy of the ML-based clone detectors via the \rev{DRLSG-guided} lightweight code transformations}. \rev{Meanwhile, we further evaluate the effectiveness of \ourtool in improving the robustness of ML-based clone detectors via adversarial training. In the best case, adversarial training improved the f-measure of the model by 43.9\%.  We find among the clone pairs provided by \DRLSG, 13.2\% and 15.8\% can still successfully escape from the detection of \Astnn and \TBCCD after adversarial training.}

 To summarize, we made the following contributions:  
  \begin{itemize}[leftmargin=*,topsep=0pt,itemsep=0ex] 
  	  \item 
  	The design and implementation of \ourtool, \rev{which supports various  heuristic strategies 
  	  to guide the   code transformations} and generate
  	  semantic-preserving code clones to challenge the clone detectors.  
  	
  \item 
	    \xyx{The proposal of a new DRL-based strategy that dispatches multiple lightweight yet effective code transformation operators at different program locations}.  

  \item  
    The evaluation of \ourtool upon the  ML-based detectors (\Astnn, \TBCCD, and  \TokenLSTM ) and the traditional detectors (\NICARD and \Deckard). \rev{Results show \ourtool can sharply decrease the detection accuracy of these detectors.}    
  
  \item  
     \xyx{A further adversarial training of ML-based detectors using the clones produced by \ourtool, which can improve the robustness of the assessed ML-based detectors.} \rev{Notably,  the DRL-based strategy is most effective in lowering the detection accuracy of the adversarial trained detectors. }
\end{itemize}

%% file: motivation.tex
\input{motivationExample}
\section{Preliminaries}
\label{sec:mtv}
\subsection{Types of Clones and Clone Detection Methods}
\label{sec:mtv:clonetypes}

\begin{myDef}
(Clone Types) {According to ~\cite{BellonKAKM07}, code clones are generally categorized into four types, namely Type I, II, III, and IV, which represent the degrees of code similarity between an original code piece and a new one, respectively.} 
\end{myDef}
 
\begin{itemize}[leftmargin=*,topsep=0pt,itemsep=0ex] 
\item {\textbf{Type I}:} The two code snippets are mostly identical except for 
the comments, indentation, and layout.
\item {\textbf{Type II}:} The differences between two code snippets are limited in  variable names, function names, types, etc, in addition to differences due to Type I. 
\item {\textbf{Type III}:} The two code snippets have slight modifications such as changed, added, removed, or reordered statements, in addition to differences due to Type I and II.
\item {\textbf{Type IV}:} Type IV only preserves semantic similarity. Thus, the two code snippets may have similar functionality but different structural patterns.
\end{itemize}

Type I--III code clones are usually referred to as \emph{syntactic} clone, which copies the code fragments and retains a large body of textual similarity. By 
In contrast, if the copied code merely exposes functional similarity, such as cloning 
presents a Type IV clone, a so-called \emph{semantic} clone.

Over years, various clone detectors have appeared in the literature~\cite{roy2009comparison,walker2020open}.
Typically, they convert software code under detection into specific representations 
and develop methods to distinguish the similarities against 
the generated representations. Before the emergence of ML-based detectors, traditional detectors (e.g., \CCAligner \cite{WangSWXR18}, \Deckard~\cite{JiangMSG07}, \CCGraph~\cite{Zou2021}) mainly rely on \emph{Token}-, \emph{AST}-, or \emph{PDG}-representation to measure the code similarity. Usually, they can
successfully detect Type I--III syntactic clones, but fail to match Type IV semantic clones. 
\Astnn~\cite{Zhang2019} splits the whole \emph{AST} into small statement trees for the finger-grained encoding of the program lexical and syntactic information. \FCDector~\cite{FangLS0S20} adopts a joint code representation of syntactic and semantic features generated from fusion embedding.
\TBCCD~\cite{YuLCLXW19} applies a tree-based convolution for semantic clone detection, which leverages both the structural information from \emph{AST} and the lexical information from code tokens. 
{ 
These ML-based detectors all achieve a stunningly high detection accuracy on the experimental datasets, \rev{often with the detection accuracy of more than 95\%}~\cite{FangLS0S20,Zhang2019,YuLCLXW19}.}


\subsection{An Example of Clone Detection}
\label{sec:mtv:example}
Fig.~\ref{fig:motiv_ex} displays a code snippet $S$ and the Type IV and Mutated of clones on $S$. 
Code $S$ come from the 
 OJClone~\cite{MouLZWJ16} database. 
{It is to find the minimum number of notes of different currency denominations
(e.g., \codeff{\{100,50,20,10,5,1\}}) that sum up to the given amount (i.e., the input variable
\codeff{n})}. 
All the clones in Fig.~\ref{fig:motiv_ex} are solutions to the same problem, but
with different levels of syntactic changes. Specifically, 
{Type IV instance is more compact --- it uses division instead of continuous subtraction to obtain integer quotients at line 11, and finish the calculation and printing in one for a loop}. 

%
For this  motivating example, the well-trained \Astnn, \TokenLSTM, and \TBCCD  can all correctly match 
the Type IV 
code (ref. code snippets \#2 in Fig.\ref{fig:motiv_ex}) with the original code (ref. code \#1 in Fig.\ref{fig:motiv_ex}) --- ML-based detector success in  identifying the syntactic and semantic clones. 
By far,  
these ML-based detectors seem to be quite effective,  
\emph{but can they thoroughly address the semantic clone detection problem}?


\subsection{\xyx{Motivation}}
\label{sec:mtv:limitation}

With the above question, we manually create a clone code by applying several
lightweight code transformations. \rev{Given the code snippet \#1 in Fig.~\ref{fig:motiv_ex}}, after changing the variable name \codeff{k} {at line 4}, splitting large constants into smaller ones {at line 3}, swapping 
the code order {at lines 3,4}, removing part of the code comments {at line 5}, and  converting 
a \textit{for-loop} to  \textit{while-loop} {at lines 7,8,9,12},  
we generate the  code snippet \#3 in Fig.~\ref{fig:motiv_ex}).


Then, we construct two clone instances as  \{\#1, \#3\} and \{\#2, \#3\}  from pairing the new code \#3 with existing code pieces in~Fig.~\ref{fig:motiv_ex}.
Now, we run \Astnn, \TokenLSTM, and \TBCCD on these new clone instances. 
We are curious to see whether a few uncomplicated code variants could lead to 
some different findings.
Surprisingly, experimental results show that all three detectors failed to classify all two code pairs to be code clones. 
This interesting result implies that {some cheap code-level changes could indeed nullify the DNNs in these detectors, without using 	dedicated adversarial samples or heavy code obfuscation (e.g.,  \xyx{obfuscation based on encoding~\cite{kovacheva2013efficient,fukushima2008analysis,xu2017secure} and CFG-flattening}~\cite{laszlo2009obfuscating,wang2000software})}. 
This observation motivates us to investigate the following questions to facilitate the automation  of adversarial code clone generation: 

\begin{itemize}[leftmargin=*,topsep=0pt,itemsep=0ex]
	
	\item \xyx{Can code clones from lightweight  semantic-preserving  transformations steadily invalidate the detection of ML-based detectors?}
	
	\item \rev{What kind of \emph{transformation strategy} is needed to guide the effective searching process of combining these semantic-preserving transformations}? 

	%
	\item  Can we leverage the new clone instances from the semantic-preserving transformations to improve the ML-based detectors and would the improvements be explicitly beneficial? 
	
	
\end{itemize}


To answer these questions, we propose and implement the clone generation framework \ourtool.

%% file: motivationExample.tex
\lstset{
		frame=single, 
		language=c++,
		backgroundcolor=\color{white},
		extendedchars=true,
		basicstyle=\scriptsize,
		numbers=left,
		numberstyle=\tiny\color{gray},
		numberstyle=\color{gray},
		keywordstyle=\color{blue}, 
		numbersep=5pt,
		xleftmargin=.25in,
		breaklines=true,
		captionpos=b,
		keepspaces=true,
		stepnumber=1,
		aboveskip=2mm,
		belowskip=2mm
}

\begin{figure*}[t]
	\centering
	\begin{minipage}[t]{0.27\textwidth}
		\centering
		\begin{lstlisting}  
// #1 Original code
int main(){
	int a[6]={100,50,20,10,5,1},b[6];
	int n,i,k;
	scanf("%d",&n);//input
	for(i=0; i<6; i++){
		b[i]=0;}
	for(k=0; k<6; k++){
		for(i=0; n>=a[k]; i++){
			n=n-a[k];
			b[k]+=1;}}
	for(i=0; i<6; i++){
		printf("%d\n",b[i]);}
		return 0; 	}
		\end{lstlisting} 
	\end{minipage}
	\begin{minipage}[t]{0.25\textwidth}
		\centering  
		 \begin{lstlisting} 
// #2 Type IV clone code
int main()
{
	int n,a[6],i;
	int b[6]={100,50,20,10,5,1};
	scanf("%d",&n);
	for(i=0;i<6;i++)
	{
		a[i]=n/b[i];
		n=n-a[i]*b[i];
		printf("%d\n",a[i]);
	}
	return 0;
}
		\end{lstlisting}
	\end{minipage}
	\begin{minipage}[t]{0.45\textwidth}
		\centering 
		\begin{lstlisting} 
// #3 Code after equivalent transformation
int main(){
	int n, i, k1;
	int a[(3 + 2 + 1)] = {100, 50, 20, (5 + 2.5 + 2.5), (2.5 + 1.25 + 1.25), (0.5 + 0.25 + 0.25)};
	int b[6];scanf("%d", &n);{  i = (0 + 0 + 0);
		while (i < 6) {b[i] = (0 + 0 + 0); 
			i = i + 1; }; }
	{k1 = 0;while(k1 < 6) {{i=0;
		while(a[k1] <= n){i++;n = n - a[k1];
			b[k1] += 1;};}k1 = k1 + 1;};
	}{i = 0;while(i < 6) {
			printf("%d\n", b[i]);i++;};
} return 0;}
		\end{lstlisting} 
	\end{minipage}
\vspace{-4mm}
\caption{Motivating example of code clone at different levels. All the clone instances are of the same semantics.} 
\label{fig:motiv_ex}
\vspace{-4mm}
\end{figure*}


%% file: approach.tex
\input{systemOverview}

\section{{\ourtool: An Overview}}
\label{sec:overview}
In this section, we describe the overall workflow for evaluating ML-based clone detectors, explain the three technical challenges we must address.  
 
\subsection{The Overview of \ourtool}\label{sec:overview:sketch}

Fig.~\ref{fig:system-overview} illustrates the \xyx{overview of \ourtool}.
In general, \ourtool consists of two phases: the \emph{clone generation}
phase and the \emph{detector evaluation} phase. The former takes a source code snippet as input, performs code equivalence transformations, and outputs the mutated code snippets. The latter evaluation phase feeds the original code snippets together with the newly generated ones to a set of state-of-the-art clone detectors and outputs whether the new clone pairs could \xyx{be detected by} these detectors.

For example, given  
code \#1  in Fig.~\ref{fig:motiv_ex},  
the steps of producing  code snippet \#3 by \ourtool are as follows. 
First, 
it extracts 
features of code snippet \#1 and searches for the locations where code transformation could be applied (see \S~\ref{sec:FCE:featureCountExtr}). Second, it adopts the predefined 15 atomic transformation operators to make sure that all the performed changes preserve the semantics of the original code (see \S~\ref{subsec:theMutOp}). Third, \ourtool also adopts a certain 
transformation strategy that properly adjusts the probability of activating an individual transformation operator (see \S~\ref{sec:encoding}).  
{Finally, by applying the  transformation operators in the order determined by the searching strategy, new code that is more likely to escape from existing detectors can be generated (see \S~\ref{sec:FCE:Code_Random_Mutation}).}
Till now, the  {clone generation}
phase is finished. 
\vspace{-1mm}
 
\subsection{Technical Challenges}
\label{sec:overview:challenge}
 	
While the high-level idea appears to be straightforward, there are three 
challenges 
to address for realizing \ourtool:

\begin{itemize}[leftmargin=*,topsep=0pt,itemsep=0ex]
\item  \noindent\textbf{Transformation Operator.} 
In this work, we require equivalence transformations, which slightly modify the code syntactic but leave the 
code semantics intact. Therefore, the  operators in this study are different from \emph{mutation operators} in \emph{mutation testing}~\cite{mutation_testing}.  
Mutation testing mutates source code for inserting small faults into programs to measure the effectiveness of a test suite. 
So mutators in mutation testing usually change code semantics during generating new code. However, we aim to achieve semantic equivalence transformations. Towards this goal, we adopt many transformation operators proposed in~\cite{lava} and~\cite{RoyC09}. In total, \ourtool supports 15 unique transformation operators, including variable renaming, changing syntax structures with equivalent semantic, adding junk code, deleting irrelevant code, 
reordering independent code, etc. Details and examples can be found in \S~\ref{subsec:theMutOp}.

\item \noindent\textbf{Encoding Schema.} 
Before applying transformation operators, it is required to localize potential code places applicable to  proper operators. For example, we must locate all the \codeff{for-loop} in code before we can decide whether to execute the operator of Op2 (see Op2 in Tab.~\ref{tab:Mutation_Example}). 
Thus, the code fragments that satisfy the transformation conditions are abstracted to execute the encoding. 
The abstraction needs to be fast and scalable as 
we target the large  codebase in evaluation. To address this issue, \ourtool develops a compact bitvector-based representation for encoding the  search space of changing a program function. Specifically, suppose a function 
owns $n_v$ variables (i.e., function name, function arguments, local variables and used global variables), $n_f$ \codeff{for-loop} statements, $n_w$ 
\codeff{while-loop} statements, $n_{do}$ \codeff{do-while} statements, $n_{ie}$ 
\codeff{if-elseif} statements, $n_{i}$ \codeff{if-else} statements, $n_{s}$ 
\codeff{switch} statements, $n_r$ relational expressions, $n_u$ unary operations 
(e.g., \codeff{i++}), $n_{sc}$  self-changing operations (e.g., \codeff{i+=1}), 
$n_c$ constant values, $n_d$ times of variable definition, $n_b$  code blocks 
($\{...\}$), $n_{is}$ blocks of isomorphic statements that have no dependency 
in the control flow, and $n_{p}$ \codeff{print} and \codeff{comment} statements. The 
length of the encoded bit vectors would be the sum of these values:
\begin{multline}\label{equation:encode}
\small 
l_b = sum(n_v,n_f,n_w,n_{do},n_{ie},n_{i},n_{s},\\n_r,n_u,n_{sc},n_c,n_d,n_b,n_{is},n_{p})
\end{multline}

During the generation phase, if a binary bit is enabled 
to be $1$, \ourtool will apply the corresponding transformation operator to the code location. However, 
it is still a difficult problem to have an optimal strategy of applying operators that match specific 
code structure for an equivalence transformation --- \xyx{generating semantic clones with various optimization goals}. Thus, we need a \emph{transformation 
strategy}.

\item \noindent\textbf{Transformation Strategy.} As mentioned above, after defining 
transformation operators and encoding the search space, 
there is another challenge: how should we decide the chance of being enabled (i.e., $1$) for each bit after encoding, to maximize the diversity of the generated semantic clones --- \rev{to challenge the existing ML-based clone detectors}? 
Generally, it appears to be a combinatorial optimization problem with a huge search space (for example, it is $2^{45 }$ for the code snippet \#1 in Fig.~\ref{fig:motiv_ex}) that could
be addressed by \xyx{commonly-used} heuristic strategies, such as  Random-Search (RS),
Genetic Algorithm (GA)~\cite{Back96}, Markov chain Monte Carlo (MCMC)~\cite{Richey10}, and the recently popular Deep Reinforcement Learning (DRL), etc. These strategies usually need some goals or objective functions (e.g., the fitness function in GA).   
\rev{Regarding this challenge, in \S~\ref{sec:encoding}, for the commonly-used strategies (RS, GA and MCMC), we  implement these  strategies to guide the transformation based on some existent studies ~\cite{liu2017stochastic,devore2020mossad,ren2021unleashing}. However, it is not straightforward to leverage DRL for this problem, for which we propose the \DRLSG model 
in \S~\ref{sec:DRLSG}.
}

\end{itemize}

%% file: systemOverview.tex
\begin{figure}[t]
	\centering
	\includegraphics[width=\linewidth]{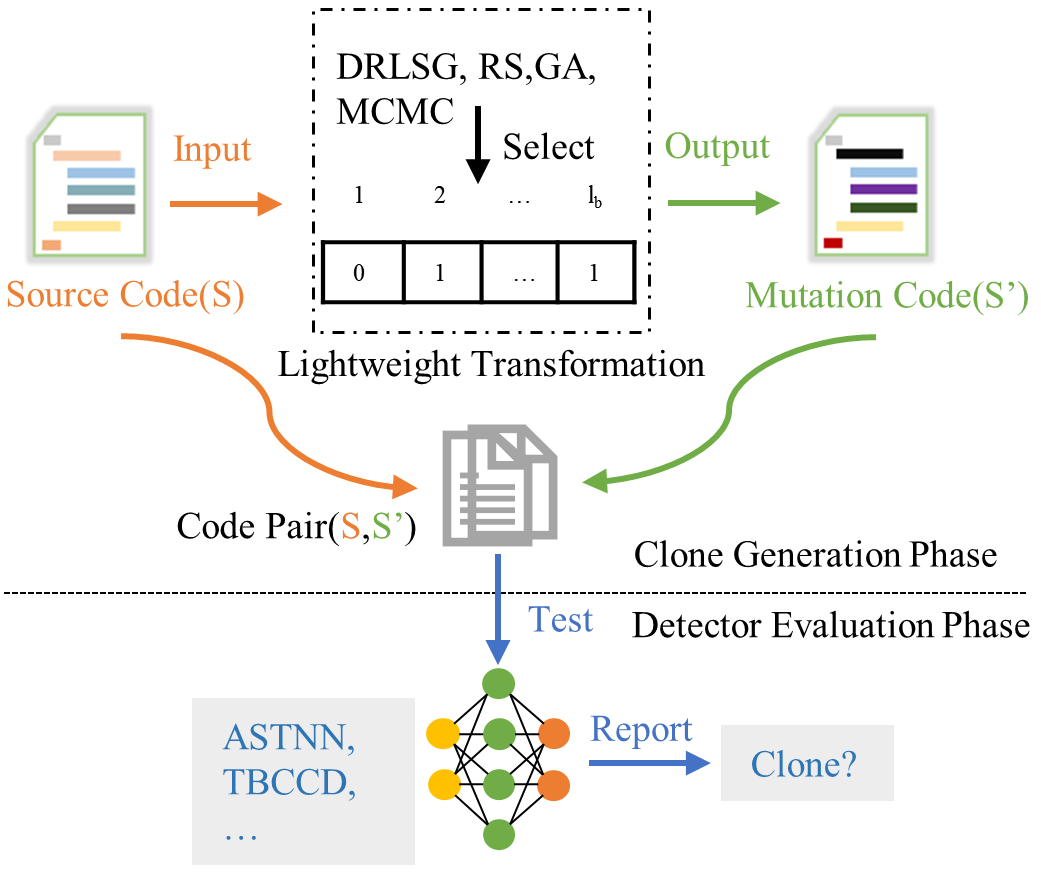}
 	\vspace{-4mm}
	\caption{\ourtool system overview} 
	\vspace{-4mm}
	\label{fig:system-overview}
\end{figure}

%% file: operator.tex
 \section{{\ourtool: Technical Approach}}
 \label{sec:prelim}

\input{operatorMutationExample}

In this section, we elaborate on major steps of \ourtool.

\subsection{Transformation Operator}
\label{subsec:theMutOp}
\ourtool transforms the given program source code by a set of atomic operations. These code transformations must not change the semantic of a program. 
Thus, we formally define these operations as \emph{equivalence transformations}.  

\begin{myDef}
Equivalence Transformation.  
{Let  $ C\stackrel{\tau}{\longrightarrow} C^\prime $  be a transformation ($\tau$) of a code snippet $C$ into a code snippet $C^\prime$ based on the combinations of a set of atomic code transformation operators $\mathcal{S}_o$, so $C$ and $C^\prime$ form a pair of semantic clones.}
\end{myDef}

Tab.~\ref{tab:Mutation_Example} summarizes all the atomic transformation operators in \ourtool. 
The first column lists the operator names, the second column briefly describes how each atomic operation takes effect, the third column shows a simple code example before equivalence transformation and the last column gives the transformed code after the transformation. Fig.~\ref{tab:SuEq} Representation of successive formulations of code transformations.
We have implemented these operators based on \textsc{Txl}\footnote{Txl: a programming language that converts input to output by a set of conversion rules, \url{https://www.txl.ca/txl-abouttxl.html}.}. 

{Transformations based on these  15 operators defined in Tab.~\ref{tab:Mutation_Example} is \emph{lightweight}, as only lexical analysis in \textsc{Txl} is adopted. Since there are no complicated operations on \emph{CFG}, \emph{PDG}, or call graph (\emph{CG}), the transformation only works at the syntax level. Meanwhile, the transformation manages to make \emph{semantic equivalence} --- all the operators guarantee to have the same semantics before and after the transformation. Even for Op4 that changes a \codeff{do-while-loop} statement into a \codeff{while-loop}, the body statement inside the 
\codeff{do-while-loop} will be executed once before entering the transformed \codeff{while-loop}. For Op14, we only reorder the statements without any data- or control-dependency, \xyx{as such analysis is fast and easy.} For Op15, we only delete comments or the statements that print debugging hints or intermediate results. For the rest of the operators, it is straightforward that these operators would not change code semantics (\rev{see operational semantics defined for these operators in Fig.~\ref{tab:SuEq}}). }

  \input{fomula}

\subsection{Encoding} 
\label{sec:FCE:featureCountExtr}

Before code transformation,  it is required to identify  {\emph{syntactic features}}, the places of  code fragments applicable to an atomic operation in Tab.~\ref{tab:Mutation_Example}. We use the term \emph{feature count} to represent the number of code 
fragments on which \ourtool performs the corresponding atomic operations.   
Tab.~\ref{tab:mutationfeatureencoding} illustrates the feature and its feature count of code \#1 in Fig.~\ref{fig:motiv_ex}. The first and second column corresponds to the atomic operations and the denoting variables in Tab.~\ref{tab:Mutation_Example} and the third column corresponds to their values. The last column lists the bitvector after encoding code \#1 in Fig.~\ref{fig:motiv_ex}.
For example, if there are $4$ \codeff{for-loop} in a source code snippet, then the
feature count of the syntactic feature \codeff{for-loop} is $n_f = 4$.  
In the third column, the value $0$ means that there are no places applicable to the corresponding atomic 
operation.  For example, no \codeff{while-loop} in code \#1, and hence its value $n_w$ is $0$. There are $5$ variables
\{\codeff{a, b, n, i, k}\} in code \#1, so the feature count for $n_v$ is $5$. Note that the \codeff{main} function cannot be renamed, so it is excluded from the feature
count of the feature $n_v$. \xyx{Hence, for code snippet \#1 in Fig.~\ref{fig:motiv_ex}, its bit-vector length $l_b$ is $45$ (the sum of column \lq\lq{}\textbf{value}\rq\rq{} in Tab.~\ref{tab:mutationfeatureencoding}).}

After we encode the transformation search space using the bit vector, the length of the bit vector equals the feature count in the second column, where an atomic operation would take effect on the features (i.e., applicable places) for bit value $1$, and no effect for $0$. The symbol "n.a."  means there are no available features that match the corresponding atomic operation, and the feature count must be $0$. We explain how to fill the bit vector 
with various strategies in the next section. 

\input{mutationFeatureEncoding}

\subsection{Transformation  Strategy}  
\label{sec:encoding}	
After the encoding step, \ourtool needs a transformation strategy to \xyx{generate the semantic clones of diversity.}
 
As discussed earlier, the search problem of finding an optimal sequence of applying the atomic operations can be treated as a combinatorial optimization problem. Thus, we may achieve different solutions by exploring existing optimization algorithms. In \ourtool, we have already implemented three commonly-used heuristic strategies, namely
Random-Search (RS), Genetic Algorithm (GA), Markov Chain Monte Carlo (MCMC). Details of each strategy are as follows:

\begin{itemize}[leftmargin=*,topsep=0pt,itemsep=0ex] 
	
	\item[] \textbf{Random-Search}: The RS strategy fills the bits in a bit vector with random values 0 or 1, with equal probability. Hence, this strategy does not take into account the source code structure under detection and the features employed by different clone detectors.  \xyx{Hence, the RS strategy favors generating semantic clones whose equivalent code changes are attributed to randomness~\cite{2019Weighted}}.
	
	\item[] \textbf{Genetic Algorithm}: The GA~\cite{michalewicz2013genetic} strategy computes the {similarity} between the original 
	code and the generated new code after each random generation. {The idea is to generate the code that exhibits the most textual differences from the original code.}
	Here we convert
	both code versions into strings and then calculate the string editing distance~\cite{navarro2001guided} between them to obtain the code similarity. An editing distance reflects how many modification steps it needs to convert one string into another, and it is used as the \emph{fitness function} in GA. \xyx{Hence, the GA strategy favors generating semantic clones with big gaps (large editing distance)~\cite{devore2020mossad,ren2021unleashing}}.

	\item[] \textbf{Markov Chain Monte Carlo}: The MCMC strategy uses the \emph{n-gram} algorithm to calculate the
	probability that determines how likely a sub-sequence follows its prefix. It is observed that software programs have probabilistic natures, which assign probabilities to different sequences
	of words~\cite{Hindle2012,liu2017stochastic}. 
	{For a given sequence of code snippet $s=s_{1} s_{2} \cdots s_{n}$, we define $P\left(s_{n} \mid s_{n-k+1}, \cdots, s_{n-1}\right)$ to approximate  the probability of  code statement $s_{n}$ occurring after  statements $s_{n-k+1}$ to $s_{n-1}$.} \xyx{Inspired by~\cite{manning1999foundations}},  we also adopt the indicator \emph{perplexity}~\cite{manning1999foundations}, to measure the probability magnitude of code occurrence. According to n-gram ($n = k$), the formula corresponds to:
	
	\begin{equation*}
		H_{\mathcal{M}}(s)=-\frac{1}{n} \sum_{1}^{n} \log p_{\mathcal{M}}\left(s_{i} \mid s_{i-k+1} \cdots s_{i-1}\right)(\operatorname{\tiny Perplexity}) 
	\end{equation*}
	
    \rev{According to the above formula, this strategy guides code transformations through several iterations and outputs the generated clones satisfying the above perplexity condition. Hence, the MCMC strategy favors generating semantic clones hard to understand (being high perplexing)~\cite{liu2017stochastic}.}
\end{itemize}

\rev{Though the above three strategies are in general effective, they would not take into consideration the feedback (e.g., detection results) of the assessed clone detectors. To mitigate this limitation, we specially design and implement a DRL-based method, named \DRLSG, to interactively incorporate the detectors' results for generating further complicated semantic clones (see \S~\ref{sec:DRLSG}). }

 \subsection{Clone Generation}  
 \label{sec:FCE:Code_Random_Mutation}
 
As shown in Fig.~\ref{fig:system-overview}, \ourtool performs a lightweight transformation phase guided by different strategies. 
For example, there are four \codeff{for-loop} in the original 
code of Fig.~\ref{fig:motiv_ex}. The last column of Tab.~\ref{tab:mutationfeatureencoding} shows the bit vectors
from the RS strategy. As stated in \S~\ref{sec:FCE:featureCountExtr}, 
Bit value 1 in a bit vector means that 
\ourtool must execute the code mutation operation at the corresponding code fragment, 
while bit value 0 means \ourtool should keep the original code piece unchanged.  
Since all bits for operator, Op2 have the value of 
$1$, \ourtool transforms the four \codeff{for-loop} into semantics preserving 
\codeff{while-loop}, as shown by code \#3 in Fig.~\ref{fig:motiv_ex}. 
Due to the randomness of these strategies, we could obtain a large number of code variants (e.g.,  code \#3 to code \#1 in Fig.~\ref{fig:motiv_ex}). \rev{For the strategies other than RS (i.e., GA, MCMC, and \DRLSG), not all code variants will be retained --- for effectiveness, only those satisfying certain optimization goals (e.g., perplexity)  will be retained.} 

Notably, our atomic operations are lightweight: according to our experiments (as described in \S~\ref{sec:evaluation}), the transformation operations can be completed within a relatively short time. 

%% file: operatorMutationExample.tex
\begin{table*}[t]
\scriptsize
\caption{The descriptions and examples of 15 atomic transformation operators in \ourtool}
\newcommand{\tabincell}[2]{\begin{tabular}{@{}#1@{}}#2\end{tabular}}  
  \centering
  \begin{tabular} {p{3cm}p{5cm}p{4.2cm}p{4.2cm}}
\toprule

\textbf{Transformation Operator} & \tabincell{c}{\textbf{Description}} &  \tabincell{c}{\textbf{Original}} & \tabincell{c}{\textbf{Changed}} \\
\midrule
Op1-ChRename & Function name and variable name renaming. &  \tabincell{l}{int i;} & \tabincell{l}{int i1;} 
 \\
\hline
Op2-ChFor & The for-loop is transformed into a while-loop. & \tabincell{l}{    for(i=0;i<10;i++)\{\\ \quad  BodyA\\\}  } &  \tabincell{l}{ i=0; while(i<10)\{ \\ \quad BodyA \\ \quad i++; \} }
 \\
\hline
Op3-ChWhile &The while-loop is transformed into  a for-loop. & \tabincell{l}{while(i<10)\{ \\ \quad BodyA \} } &\tabincell{l}{    for(;i<10;)\{\\ \quad BodyA\}  } \\
\hline
Op4-ChDo & The do-loop is transformed into a while-loop. & \tabincell{l}{do\{\\ \quad BodyA \\\}while(i<10);} & \tabincell{l}{BodyA\\while(i<10)\{ \\\quad BodyA\}
}\\
\hline
 Op5-ChIfElseIF  & Transformation of if elseif to if else. & \tabincell{l}{ if(grad<60)  BodyA\\ else if(grad<80)   BodyB \\ else  BodyC } & \tabincell{l}{if(grad<60) BodyA \\ else\{    if(grad<80)   BodyB \\  \quad else  BodyC \}}\\
\hline
Op6-ChIf & Transformation of if else to if elseif. & \tabincell{l}{if(grad<60) BodyA \\ else\{ if(grad<80)   BodyB \\  \quad else  BodyC \}} &\tabincell{l}{ if(grad<60)  BodyA\\ else if(grad<80)   BodyB \\ else  BodyC }  \\
\hline
Op7-ChSwitch &  Transformation of the  Switch statement to the  if elseif statement. & \tabincell{l}{switch(a)\{  case 60: BodyA \\ \quad case 70: BodyB\\ \quad default: BodyC \}} & \tabincell{l}{if(a==60) BodyA\\ else if(a==70) BodyB\\ else BodyC }\\
\hline
Op8-ChRelation &  Transformation of relational expressions. & \tabincell{l}{a<b} & \tabincell{l}{b>a}\\
\hline
Op9-ChUnary  &  Modifications to unary operations. & \tabincell{l}{i++;} & \tabincell{l}{i=i+1; }\\
\hline
Op10-ChIncrement &  Modifications to incremental operations. & \tabincell{l}{i+=1;} & \tabincell{l}{i=i+1;}\\
\hline
Op11-ChConstant & Modifying constants. & \tabincell{l}{8} & \tabincell{l}{(a-b)} //8=a-b\\
\hline
Op12-ChDefine & Modifications to variable definitions. & \tabincell{l}{int b=0;} & \tabincell{l}{int b;b=0;}\\
\hline
Op13-ChAddJunk &  Adding junk codes. & \tabincell{l}{if(a)\{\\ \quad  BodyA\}} &\tabincell{l}{if(a) \{    BodyA \\ \quad if(0) return 0;  \}}\\
\hline
Op14-ChExchange & Change the order of the statements in a block without data and control dependency. & \tabincell{l}{a=b+10;\\c=d+10;} & \tabincell{l}{c=d+10;\\a=b+10;}\\
\hline
Op15-ChDeleteComments &  Deleting   statements  that   print     debugging hints and comment.
& \tabincell{l}{printf("test");\\//comments} & \tabincell{l}{\sout{//printf("test");}\\\sout{//comments}}\\
\bottomrule
\end{tabular}

 \vspace{-2mm}
 \label{tab:Mutation_Example}

\end{table*}

%% file: fomula.tex
\begin{figure*}[t]
   \small
	\centering
	\vspace{-3mm}
	\begin{minipage}[!h]{0.99\linewidth}
		\begin{align*}
			\begin{aligned}
				&Symbols:\  \rightarrow \ means\ syntax\ transformation,\ \Downarrow \ means\ evaluation\ operation,\ Change\ function\ is\ a\ bijection\ function,\ \\ & Dependency\ is\ a\ data\ and\ control\ dependency\ statements\ set\ in\ a\ block,\ Comments\ is\ a\  comment\ statement\ set,\ Printf\\ & is\ a\ printf\ call\ statements\ set\ with\ constant\ arguments.
			\end{aligned}	
		\end{align*}
	\end{minipage}

		\begin{minipage}[t]{0.44\textwidth}
			\begin{align*}
				\begin{aligned}
					E \rightarrow E' &\quad S \rightarrow S'\\
					\hline
					do \ S\ \quad while(E) \quad \rightarrow &\quad S'; while(E')\ S' 
				\end{aligned}
				\tag{Op4-ChDo}
			\end{align*}
		\end{minipage}
		\begin{minipage}[t]{0.55\textwidth}
			\begin{align*}
				\begin{aligned}
					S_1 \rightarrow S_1' \quad E \rightarrow E'&\quad S_2 \rightarrow S_2' \quad S_3 \rightarrow S_3' \\
					\hline for(S_1; E;S_2)\ S_3 \rightarrow & \{S_1'\}; \quad while(E') \{S_3'; S_2'\}
				\end{aligned}
				\tag{Op2-ChFor}
			\end{align*}
		\end{minipage}

		\begin{minipage}[t]{0.44\textwidth}
			\begin{align*}
				\begin{aligned}
					E &\rightarrow E'\\
					\hline
					E\ \texttt{+=}\ 1 \quad \quad \rightarrow &\quad E'\ \quad \texttt{=}\ \quad E'\ \texttt{+}\ 1
				\end{aligned}
				\tag{Op10-ChIncrement}
			\end{align*}
		\end{minipage}
		\begin{minipage}[t]{0.55\textwidth}	
			\begin{align*}
				\begin{aligned}
					E \rightarrow E' &\quad S \rightarrow S'\\
					\hline
					if (E)\ \{S\} \rightarrow if(E')\quad&\{S'; if(0)\ return\ 0;\}
				\end{aligned}
				\tag{Op13-ChAddJunk}
			\end{align*}
		\end{minipage}
		
		\begin{minipage}[t]{0.42\linewidth}
			\begin{align*}
				\begin{aligned}
					S \rightarrow S' &\quad E \rightarrow E'\\
					\hline
					while(E)\ \quad S\quad &\rightarrow \quad for(;E';)\ S'
				\end{aligned}
				\tag{Op3-ChWhile}
			\end{align*}
		\end{minipage}
		\begin{minipage}[t]{0.57\textwidth}
			\begin{align*}
				\begin{aligned}
					S\in &Comments \lor 
					F \in Printf
					\\
					\hline
					S &\rightarrow \{\} 
					\quad
					F(E) \rightarrow \{\}
				\end{aligned}\tag{Op15-ChDeleteComments}
			\end{align*}
		\end{minipage}

		\begin{minipage}[t]{0.42\textwidth}	
			\begin{align*}
				\begin{aligned}
					E \rightarrow E' &\quad id \rightarrow id' \\
					\hline
					int\ id\ \texttt{=}\ E \quad \rightarrow &\quad int\ id';\ id'\ \texttt{=}\ E'
				\end{aligned}
				\tag{Op12-ChDefine}
			\end{align*}
		\end{minipage}
		\begin{minipage}[t]{0.57\textwidth}		
			\begin{align*}
				\begin{aligned}
					N_1 \in [10..1000] \quad N_1,&N_2,N_3 \in\mathbb{Z} \quad\quad N_2 \texttt{-}N_1\Downarrow N_3\\
					\hline
					N_3\rightarrow (&N_2 \texttt{-}N_1)
				\end{aligned}\tag{Op11-ChConstant}
			\end{align*}
		\end{minipage}

        \begin{minipage}[t]{0.33\textwidth}
			\centering
			\begin{align*}
				\begin{aligned}
					E_1 \rightarrow E_1' \quad E_2 \rightarrow E_2'\\
					\hline
					E_1 < E_2 \rightarrow  E_2' > E_1'
				\end{aligned}
				\tag{Op8-ChRelation}
			\end{align*}
		\end{minipage}
		\begin{minipage}[t]{0.66\textwidth}
			\begin{align*}
				\begin{aligned}
					E_1 \rightarrow E_1' \quad S_1 \rightarrow \quad S_1' \quad E_2 &\rightarrow E_2' \quad 
					S_2 \rightarrow S_2' \ \quad \quad S_3 \rightarrow S_3' \\
					\hline
					if(E_1)\ S_1\  else\ \{if (E_2)\ S_2\ else\ S_3\}  &\rightarrow if (E_1')
					S_1'\   else if (E_2')\ S_2'\  else\ S_3'
				\end{aligned}\tag{Op6-ChIf}
			\end{align*}
		\end{minipage}

		\begin{minipage}[t]{0.3\textwidth}
			\centering 
			\begin{align*}
				\begin{aligned}
					E &\rightarrow E' \\
					\hline
					E\texttt{++} \rightarrow &E'\ \texttt{=}\ \quad E'\texttt{+}1
				\end{aligned}
				\tag{Op9-ChUnary}
			\end{align*}
		\end{minipage}
		\begin{minipage}[t]{0.69\textwidth}
			\begin{align*}
				\begin{aligned}
					E_1 \rightarrow E_1' \quad S_1  \rightarrow S_1' \quad E_2  &\rightarrow E_2'\quad S_2 \rightarrow  S_2' \quad S_3 \rightarrow S_3' \\
					\hline
					if (E_1)\ S_1\   else if (E_2)\ S_2\   else\ S_3 &\rightarrow if(E_1')
					S_1'\   else\ \{if (E_2')\ S_2'\ else\ S_3'\}
				\end{aligned}\tag{Op5-ChIfElseIf}
			\end{align*}
		\end{minipage}

		\begin{minipage}[t]{0.34\textwidth}
			\begin{align*}
				\begin{aligned}
					id'\quad \in \quad  &change(id) \quad \\
					\hline
					id &\rightarrow id'
				\end{aligned}
				\tag{Op1-ChRename}
			\end{align*} 
		\end{minipage}
		\begin{minipage}[t]{0.65\textwidth}
			\begin{align*}
				\begin{aligned}
					S_1 \notin Dependency \quad S_2 \notin Dependency &\quad \quad
					S_1 \rightarrow S_1' \quad\S_2 \rightarrow S_2'\\
					\hline
					S_1; S_2 &\rightarrow S_2'; S_1'
				\end{aligned}\tag{Op14-ChExchange}
			\end{align*}
		\end{minipage}

		\begin{minipage}[t]{0.995\textwidth}	
			\begin{align*}
				\begin{aligned}
					E \rightarrow E' \quad S_1 \rightarrow S_1' \quad S_2 &\rightarrow S_2' \quad S_3 \rightarrow S_3'\\
					\hline
					switch(E) \{case\ N_1:\ S_1 \quad case\ N_2: S_2 \quad default:\ S_3 \} \rightarrow & if (E' \ \texttt{==}\ N_1)\ S_1\quad else\ if (E' \ \texttt{==}\  &N_2)\ S_2\   else\ S_3
				\end{aligned}\tag{Op7-ChSwitch}
			\end{align*}
		\end{minipage}
   \caption{Operational semantics for the atomic code transformation operators.} 
	\vspace{-3mm}
	\label{tab:SuEq}
\end{figure*}

%% file: mutationFeatureEncoding.tex
\begin{table} [!t]
\footnotesize
\centering 
\caption{Syntatic features and encoding of code \#1 in Figure~\ref{fig:motiv_ex}}
 \vspace{-2mm}
 \setlength{\tabcolsep}{2.5mm}{
\begin{tabular}{p{1.2cm}p{1.2cm}p{1.2cm}p{3.3cm}} 
\toprule \textbf{Operator}& \textbf{Count}&\textbf{value}&  \textbf{Bitvector Encoding} \\ 
\midrule 
Op1 & $n_v$& 5 &  001000    \\
Op2 & $n_f$& 4 &  1111  \\ 
Op3 & $n_w$& 0 &  n.a.  \\ 
Op4 & $n_{do}$& 0 &  n.a.  \\ 
Op5& $n_{ie}$& 0 &  n.a.  \\ 
Op6 & $n_{i}$& 0 &  n.a.  \\ 
Op7& $n_{s}$& 0 &  n.a.  \\ 
Op8& $n_r$& 4 &  0001   \\ 
Op9& $n_u$& 4 & 0110  \\
Op10& $n_{sc}$& 1 & 0   \\ 
Op11& $n_c$& 18 & 100011000000001011    \\
Op12 & $n_d$& 2 & 01    \\ 
Op13& $n_b$& 5 & 00000  \\ 
Op14& $n_{is}$& 2 & 10    \\ 
Op15& $n_{p}$& 0 &  n.a  \\ 

\bottomrule 
\end{tabular}
}

\label{tab:mutationfeatureencoding} 
\centering 
 \vspace{-4mm}
\end{table}

%% file: deepRL.tex
\section{DRLSG: Theory  and Design} 
\label{sec:DRLSG}
In this section, we introduce the background knowledge of reinforcement learning and our proposed Deep Reinforcement Learning Sequence Generation (DRLSG) strategy.

\subsection{{Deep Reinforcement Learning}} \label{sec:DRLSG:background}

\label{sec:mtv:DRL}
In recent years, deep reinforcement learning (DRL) algorithms have been used in a variety of fields~\cite{levine2016end,silver2017mastering,mnih2015human}, most notably AlphaGo\cite{silver2016mastering} to defeat the best human players in Go, and DRL has quickly become the focus of the artificial intelligence community.
This paper is based on deep reinforcement learning to generate high-quality source code for transformation sequences. Before we start presenting our approach, we briefly introduce background knowledge about DRL. The following formulation of a typical DRL process is related to the presentation given in~\cite{szepesvari2010algorithms}  



We briefly introduce the terms commonly used in DRL:
\textbf{Agent}: The role of a learner and decision-maker.
\textbf{Environment}: Everything that is composed of and interacts with something other than the agent.
\textbf{Action}: The behavioral representation of the agent body.
\textbf{State}: The information that the capable body obtains from the environment.
\textbf{Reward}: Feedback from the environment about the action.
\textbf{Strategy}: The function of the next action performed by the agent based on the state.
\textbf{on-policy}: The policy that corresponds when the agent learns and the agent interacts with the environment are the same. 
\textbf{off-policy}: The policy when the agent to be learned and the agent interacting with the environment are not the same.

In reinforcement learning, a policy indicates what action should be taken in a given state and is denoted by $\pi$. If we use deep learning techniques to do reinforcement learning, the strategy is a network. Inside the network, there is a set of parameters, and we use $\theta$ to represent the parameters of $\pi$ ~\cite{sutton2018reinforcement}. We take the environment output state ($s$) and the agent output action ($a$), and string $s$ and $a$ all together, called a Trajectory ($\tau$), as shown in the following equation:$\tau=\left\{s_{1}, a_{1}, s_{2}, a_{2}, \cdots, s_{t}, a_{t}\right\}$ .You can calculate the probability of each trajectory occurring:

\begin{equation}
	\begin{aligned}
		p_{\theta}(\tau) 
		&=p\left(s_{1}\right) \prod_{t=1}^{T} p_{\theta}\left(a_{t} \mid s_{t}\right) p\left(s_{t+1} \mid s_{t}, a_{t}\right)
	\end{aligned}
\end{equation}

The reward function determines how many points are available for said action now based on a certain action taken in a certain state. What we have to do is to adjust the parameter $\theta$ inside the actor so that the value of $\bar{R}_{\theta}=\sum_{\tau} R(\tau) p_{\theta}(\tau)$ is as large as possible. We use gradient ascent because we want it to be as large as possible. We take a gradient for $\bar{R}$, where only $p_{\theta}$ is related to $\theta$, so the gradient is placed at $p_{\theta}$.

\begin{equation}
	\theta \leftarrow \theta+\eta \nabla \bar{R}_{\theta}
\end{equation}
\begin{equation}
	\begin{aligned}
		\nabla \bar{R}_{\theta} 
		=E_{\tau \sim p_{\theta}(\tau)}\left[R(\tau) \nabla \log p_{\theta}(\tau)\right] 
	\end{aligned}
\end{equation}

Proximal policy optimization(PPO)~\cite{schulman2017proximal} is a variant of policy gradient. 
Using $\pi_{\theta}$ to collect data, we have to sample the training data again when $\theta$ is updated. So we want to go from on-policy to off-policy. We change on-policy to off-policy by importance sampling, from $\theta$ to $\theta^{\prime}$. So now the data is sampled with $\theta^{\prime}$, but the parameter to be adjusted for training is the model $\theta$.

\begin{equation}
	\nabla \bar{R}_{\theta}=E_{\tau \sim p_{\theta^{\prime}}(\tau)}\left[\frac{p_{\theta}(\tau)}{p_{\theta^{\prime}}(\tau)} R(\tau) \nabla \log p_{\theta}(\tau)\right]
\end{equation}


\subsection{DESIGN OF DRLSG} \label{sec:DRLSG:strategy}

\input{rlmodel}

As shown in Fig.~\ref{fig:drl-overview}, our \DRLSG model has two main components: agent and environment. The agent mainly consists of a neural network, and here we choose the openAI open source \textsc{PPO}\footnote{Stable Baselines3 is a set of reliable implementations of reinforcement learning algorithms in PyTorch, \url{https://stable-baselines3.readthedocs.io/en/master/modules/ppo.html}.} model as our agent.  The input to the agent is the encoding vector (state) of the code, and we train the agent to choose the corresponding transformation operator \S~\ref{subsec:theMutOp}, to maximize the reward value. The environment consists of three main parts: the code transformation according to the action chosen by the agent, the code is encoded to obtain the vector as the state of the environment, and the reward is calculated for the current decision.  Next, we describe the specific design of the DRLSG.

\subsubsection{Action Space}
In our DRLSG task, the agent is trained to select the action to be performed from the action space given a state. We use the 15  transformations designed as the action space of the model, and in our task, the goal of the model is to select a set of operations from the 15  transformations to transform the source code. These transformations are shown in Tab.~\ref{tab:Mutation_Example}, where we give descriptions of the specific operations and simple examples, and in Fig.~\ref{tab:SuEq}, where we give formal definitions of the corresponding equivalent transformations.

\subsubsection{Reward}
\label{sub:sub:reward}
The reward function is designed to guide the overall actions of the agent and reward is the key to the DRL, we need to maximize the cumulative gain of the agent. The main goal of our agent is to produce variant codes that can escape detection by Ml-based clone tools.
For this purpose, our reward function for each step is formalized as Algorithm~\ref{alg:reward}.


\begin{algorithm}[t]
  \caption{Reward Algorithm}
  \label{alg:reward}
  \footnotesize
  \KwIn{\\
  \quad\quad $action$: The current action performed by the agent.\\
  \quad\quad $code_o$: The original code.\\
  \quad\quad $code_t$: The current action gets the code snippet. \\
  \quad\quad $code_{t-1}$:The code snippet of the previous state.}
  \KwOut{\\
  \quad\quad $reward$}

  \If{!clone( $code_o$,$code_t$)}
  {
    $reward$=$nclone_R$
  }
  \Else
  {
     $reward$=-$clone_R$ * codeTextSim($code_t$,$code_{t-1}$)
  }
  \If{action==Op13}
  {
    $reward$=$reward$-\text{ $Op13\_count$ } * $Op13\_penalty$ \\
    $Op13\_count$++
  }
\end{algorithm}


As shown in Fig.~\ref{fig:drl-overview}
Our reward has two main components that make up the current action  to get the edit distance between $code_t$ and $code_{t-1}$, and the current action to get the clone detection result between $code_t$ and $code_o$. They constitute our reward, and we are considering a long-term reward. We implement a text-based Siamese network source code clone detection model, using two identical bidirectional LSTM (BiLSTM) networks in Siamese's architecture. The $code_o$ and $code_t$  are converted into text sequences and fed into BiLSTM to obtain a representation of the code. 
If the \TokenLSTM identifies the $code_t$ and the $code_o$  as non-clones, we assign a large positive reward (Line 1-3 in Algorithm~\ref{alg:reward}) and abort the current learning, indicating that we have obtained a transformed sequence of the current code.

The edit distance indicates the minimum number of single-character editing operations required to convert from one word to another. There are three types of editing operations: insert, delete, and replace. For example, for the words "kitten" and "sitting", the minimum single-character editing operations required to convert from "kitten" to "sitting" are: substitution of "s" for "k", the substitution of "i" for "e", and insertion of "g" at the end, so the edit distance between these two words is 3 . We implemented a text sequence-based code text similarity based on edit distance Code Text Similarity (codeTextSim), defined as follows:


\begin{equation*}
	\begin{aligned}
	codeTextSim=
	1-\frac{editDistance(code_t,code_{t-1})}{max(len(code_t),len(code_{t-1}))}
	\end{aligned}
\end{equation*}

where $len$ denotes the length of the code sequence, and if the edit distance between two codes is 0, then their similarity is $codeTextSim=1.0$, which means there is no difference between the two codes. If the action executed by the current state model yields $code_t$ that does not change relative to the $code_{t-1}$ of the previous state, then their $codeTextSim=0$. We give it a negative penalty of $clone_R*codeTextSim$ (Line 4-6 in Algorithm~\ref{alg:reward}). This is designed to avoid the rewards in the environment being too sparse, to help improve their learning efficiency and converge as quickly as possible~\cite{ng1999policy}. since there is no limit to the amount of garbage code that can be inserted. To control the code complexity, we use $Op13_{count}$ to count the number of times op13 is used in the current round and give an additional negative penalty $Op13_{penality}*Op13_{penality}$ (Line 7-10 in Algorithm~\ref{alg:reward}) if the action selected by the current agent is op13.
\rev{In our implementation, we set $nclone_R$, $clone_R$, and $Op13_{penalty}$ to 10, 0.5, and 0.5 empirically}.

\subsubsection{state}
In DRL state represents the current state of the environment. In our model, the transformation code obtained from the current action represents our current state, which can be represented by the encoding vector of the code. For a given source code we convert it into a sequence of texts sequence and then encode it with a bidirectional LSTM model, which is the same as the clone detection model mentioned in \S~\ref{sub:sub:reward}.

%% file: rlmodel.tex
\begin{figure}[t]
	\centering
	\includegraphics[width=\linewidth]{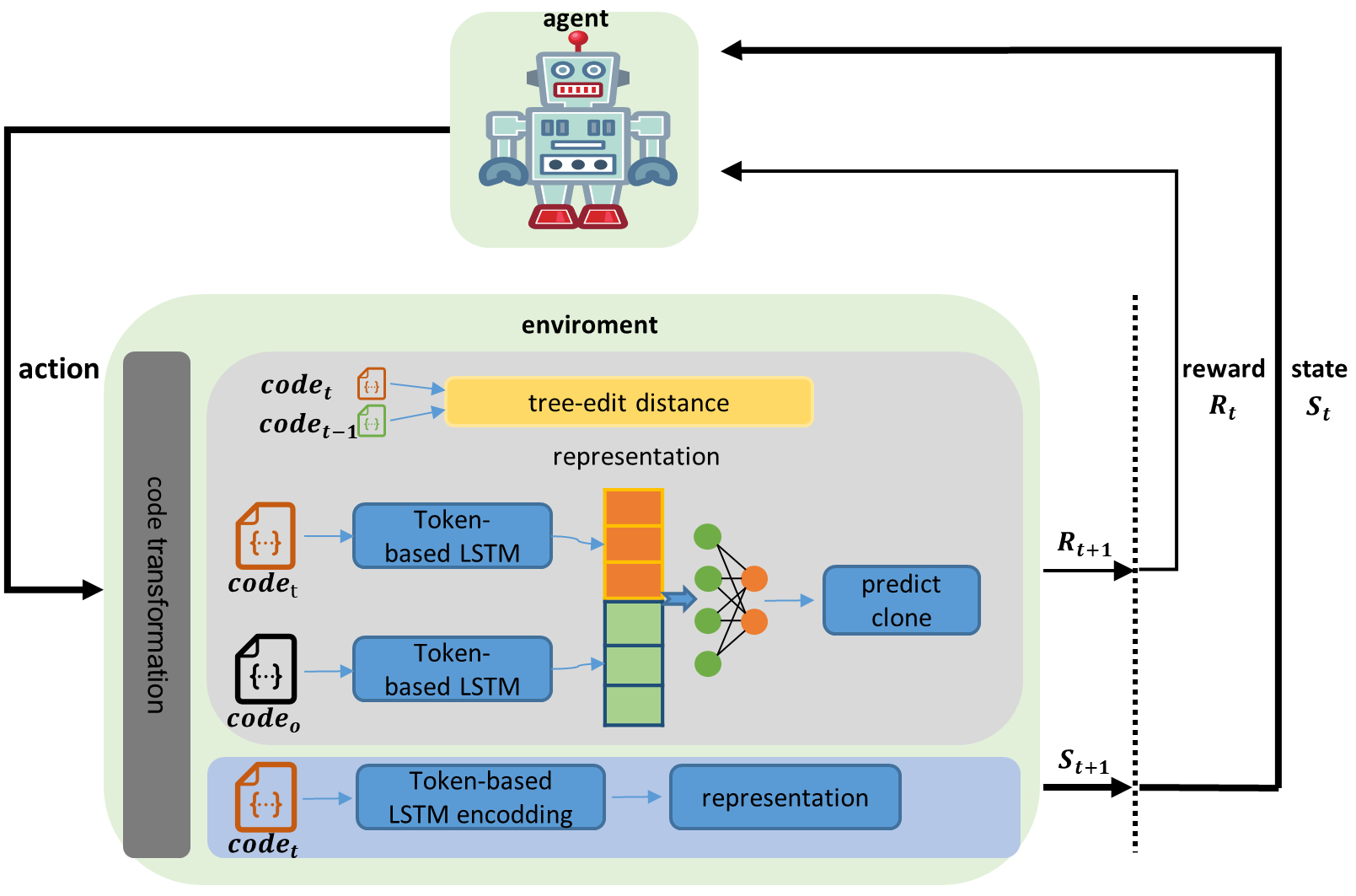}
 	\vspace{-4mm}
	\caption{Deep reinforcement learning model} 
 	\vspace{-4mm}
	\label{fig:drl-overview}
\end{figure}

%% file: evaluation.tex
 \section{Evaluation} 
 \label{sec:evaluation} 
 
In this section, we aim to investigate the following four research questions (RQs) through the experimental evaluation:
 
 \begin{itemize}[leftmargin=*,topsep=0pt,itemsep=0ex] 	
	\item \textbf{RQ1}: \rev{{How  {effective} are the different transformation strategies?}  How robustness are the existing ML-based detectors in detecting the semantic clones generated by \ourtool?}
	\item \textbf{RQ2}: \rev{How effective is our proposed \DRLSG  when detection results are available?}  \rev{{Can the ML-based detectors be  {enhanced} by adversarial training with \ourtool?}} 
	 \item \textbf{RQ3}: \rev{How  {effective} are different types of atomic transformation operators?}
	\item \textbf{RQ4}: \rev{{How  {accurate}  are the traditional  clone detectors   in detecting the semantic clones produced by \ourtool? }} 

\end{itemize}

\subsection{Implementation and experimental setup}
\label{sec:exp_setting}

\subsubsection{{Datasets for \ourtool}}
 In our evaluation, we use OJClone~\cite{MouLZWJ16}, a widely used database in source code  
 study~\cite{FangLS0S20,Zhang2019,YuLCLXW19,MouLZWJ16,wei2018positive,zhang2019find}. OJClone consists of 104 folders, each of which contains  500 solutions (clones) to the same problem.
The code from each OJClone folder to form an initial dataset $D_{I}$, and 
 apply the four strategies (RS, GA, MCMC and DRLSG) to $D_{I}$ to construct four new datasets, \rev{denoted as $D_{RS}$, $D_{GA}$, $D_{MCMC}$ and $D_{DRL}$, respectively.}

\input{clonePairsAndGenTime}

Tab.~\ref{tab:dataPairsAndGenTiem} shows the time consumed to transform 52,000 source code snippets in the OJClone dataset by four transformation strategies.
\rev{For each code snippet in the 52,000, we apply the four different strategies to generate code pairs. Notably, the RS strategy generates one code variant for each snippet, while the other three strategies would generate many clone candidates and then use the corresponding optimization goal to retain the best one. Finally, we have the same size for these generated datasets: $|D_{I}| = |D_{RS}|= |D_{GA}| = |D_{MCMC}| = |D_{DRL}|.$}

The RS strategy takes only 0.88 hours, which is the fastest. The GA and MCMC strategies need to do some code syntax analysis when transforming the code, so they are slower than the RS strategy.
The DRLSG strategy requires an ML-based encoder to encode the source code and an ML-based detector to perform similarity analysis, and only one operator is selected for each transformation, so it is much slower.

\subsubsection{{Assessed Clone Detectors}}
There are many ML-based C++ source code clone detectors, including ~\cite{FangLS0S20,zhang2019find,Zhang2019,zhao2018deepsim,zhang2019find,YuLCLXW19,wei2018positive,MouLZWJ16,wang2020modular,mou2016convolutional} etc. However, some of them are not open source, \rev{and some others have implementation issues when handling our dataset (e.g., \FCDector~\cite{FangLS0S20}) \textcolor{black}{}}. Finally, we adopt two open-source detectors \Astnn~\cite{Zhang2019} and \TBCCD~\cite{YuLCLXW19}. We also apply the \TokenLSTM~\cite{xingjian2015convolutional} model on the OJClone dataset. In the training/testing steps, we follow the guidelines provided in their GitHub pages, \rev{and use the suggested parameters in their papers~\cite{Zhang2019, YuLCLXW19} to make fair comparisons.}
\rev{Here, \TokenLSTM is a model with a text embedding size of 300 and a BiLSTM hidden layer size of 256. During the training process we divided the dataset into training, validation, and test sets according to 80\%, 10\%, and 10\%, and we used the Adam optimizer with a learning rate of 0.001 to train 20 epochs and save the model with the best F1 during the training. The above hyperparameters are empirically determined according to our experiments. }

\rev{The detailed results are shown in Tab.~\ref{tab:tokenlstmPer}, \Astnn and \TBCCD are tested by our replication. They both perform well on OJClone with $F1$ greater than 0.96. OJClone has a big difference and belongs to type IV clone, which means that the existing ML-based clone detectors have achieved good results, here the \TokenLSTM $F1$ (0.991) is the best. Tab.~\ref{tab:dlTrainTest} summarizes the time spent on setting up the three ML-based detectors. \TokenLSTM spends the least time on data pre-processing and \Astnn spends the least time on training and testing. \TBCCD takes the maximum time, about 1.38 hours, in pre-processing and  
52 hours in training and testing. }


\input{tokenLSTMPerformance}

\input{dlTrainTestTime}

\subsubsection{Baseline Transformation Strategies}
As introduced in \S \ref{sec:encoding}, the search problem can be treated as a combinatorial optimization problem. Thus, we may achieve different solutions by exploring various optimization algorithms. In this paper, we also compare our proposed  \DRLSG with the commonly-used heuristic strategies such as RS, GA, and MCMC. \rev{We have derived the parameters for GA from the related work of~\cite{devore2020mossad,ren2021unleashing,michalewicz2013genetic}. For MCMC, we have referblack to~\cite{liu2017stochastic,zhang2020generating} for the parameter setup.} 

\subsubsection{Evaluation {Metrics}}
After obtaining the four new datasets, we feed them to clone detectors under test. \rev{Notably, to evaluate the robustness of the ML models inside the detectors, the instances in our dataset have two types of labels: clone and non-clone.
For example,  in the generated dataset $D_{RS}$, it consists of 52,000 clone pairs (each one from the original $D_I$ and the other one generated by the RS strategy) and also 52,000  non-clone pairs (code pairs from different folders transformed after the RS strategy). } 

We use Precision ($P$), Recall ($R$) and F1-Measure ($F_1$) Score to measure the performance of the ML-based detector.
Let $TP$ pblackict positive class as positive, $FN$ pblackict positive class as negative,
$FP$ pblackict negative class as positive, and $TN$ pblackict negative class as negative,
we compute $P$, $R$, and $F_1$ as follows: 

\begin{equation}
P=\frac{T P}{T P+F P} \quad R=\frac{T P}{T P+F N} \quad F_{1}=\frac{2 P R}{P+R}
\end{equation}

\rev{We use Recall (R) to measure the performance of traditional tools on detecting code clones.   Recall is an evaluation indicator commonly used by these detectors (e.g.,~\cite{SajnaniSSRL16,WangSWXR18})}.


 

\subsubsection{{Environment}}
 We conduct all the experiments on an AMAX 
 computing server.  It has two 2.1GHz  24-core CPUs, four NVIDIA GeForce RTX 3090 GPUs, and 
 384G 
 memory.

 \subsection{RQ1: ML-based Detectors vs. \ourtool }
 \label{subsec:effect_clone_detection_approach}

 In this section, we answer the \textbf{RQ1}, the effectiveness of \ourtool on bypassing the ML-based clone detectors. \rev{Towards this goal, we use the initial datasets $D_I$ as the \textbf{training dataset} for three assessed detectors. Then, for each different strategy, we use the corresponding generated dataset ($D_{RS}, D_{GA}, D_{MCMC}, D_{DRL}$) as the \textbf{testing dataset}. }

\input{dlAllTestResult}

Next, we evaluate the robustness of these detectors against the code pairs generated by \ourtool.
 Tab.~\ref{tab:oringinal_recall_e} shows the accuracy of the three detectors.  Generally, these ML-based detectors have varying accuracy regarding the four transformation strategies.

\TokenLSTM directly converts the code into a sequence of texts, and then feeds the code into an LSTM neural network to obtain the embedding vector of the code, and then feeds the vector into a shallow feedforward neural network to make a binary classification judgment, 
 \TokenLSTM has the best clone detection performance on the $D_I$ with an $F_1$ of 0.991 \rev{in Tab. \ref{tab:tokenlstmPer}}. The $F_1$ of \TokenLSTM ranges from 0.502 to 0.882 for four different strategies, all of which affect \TokenLSTM compablack to the $F_1$ on the original dataset (0.991). Among the four strategies, DRLSG (0.502) is the best in bypassing \TokenLSTM detection, while RS (0.882) is the worst, indicating that random transformations are more easily detected by \TokenLSTM.

\Astnn is a neural network built upon the \emph{AST} of a program. The \emph{$F_1$} of 
\Astnn ranges from 0.530 to 0.701. 
 In all four code transformation strategies, the DRLSG strategy performs the best and the RS strategy performs the worst. The result shows that \rev{different strategies} can \rev{extensively} bypass the 
 \emph{AST}-based detection in \Astnn.

 \TBCCD uses structural \emph{AST} and token lexical
 information to generate a tree-based 
 convolutional neural network. According to Tab.~\ref{tab:oringinal_recall_e}, \TBCCD 
 has better 
 robustness than \Astnn and \TokenLSTM, with an average \emph{$F_1$} value of 0.850. 
 \TBCCD performs best against MCMC with an  \emph{$F_1$} of 0.908, 
 indicating that \TBCCD  is relatively insensitive to MCMC-guided 
  transformation. 
 The code clone pairs generated from the {GA} strategy achieve \emph{$F_1$}=0.739  on \TBCCD, followed
 by {DRLSG} (0.863) and {RS} (0.891). This result indicates that the  token information   
 increases the resilience of ML-based detectors against code transformation.   

%
%
%
 
In summary, from Tab.~\ref{tab:oringinal_recall_e} we observe that the pure \emph{Text}-based
detector (\TokenLSTM) can be easily bypassed with a high possibility. The models based on
hybrid abstractions of source code do achieve better resilience, especially the 
\TBCCD model which relies on both \emph{Token} and \emph{AST}. From the \ourtool 
perspective, the DRLSG strategy has shown the best effectiveness since it 
performs best in testing \TokenLSTM and \Astnn.

Now, we answer the \textbf{RQ1}: the code pairs generated by \ourtool can effectively bypass the detection 
of the state-of-the-art ML-based detectors, \rev{sharply dropping \emph{$F_1$} of these detectors from 90+\% to as low as 50\%-70\% in many cases}. \TokenLSTM, \xyx{being most efficient,} is the most fragile one with a low \emph{$F_1$} of 0.421,  \TBCCD has the best robustness against \ourtool, \xyx{but it relies on very costly pre-processing and training}. \xyx{Last, \Astnn seems to strike a balance in efficiency and robustness, \xyx{as clone detection is just one of the applications \Astnn supports}.}

\subsection{RQ2: Adversarial Training with \ourtool}
\label{adv_training}

In this section, we try to answer the \textbf{RQ2} with the experimental results. That is, 
whether or not augmenting the training of ML-based detectors with adversarial samples 
can defend \ourtool. The answer to RQ1 reveals the 
robustness issues of the ML-based detectors. A common idea of reinforcing such detectors 
is to fuse the spear into the shield, which means we can convert the code clone pairs 
from \ourtool to be adversarial samples to re-train the ML models.


\input{dlMutatedAddedTestResult}

It is believed that training data significantly affects the performance of deep learning 
models~\cite{zhu2016we}. In general, the more complete the training data is, the better
performance the model achieves. \rev{Towards this goal, we enhance the original dataset $D_I$ with the generated adversarial samples by four different strategies. Consequently, we use  the following datasets as the \textbf{training dataset}: {$D_I$+ 25\%$D_{RS}^\prime$+ 25\%$D_{GA}^\prime$+25\%$D_{MCMC}^\prime$+25\%$D_{DRL}^\prime$}. 
Here,  $D_{RS}^\prime$, $D_{GA}^\prime$, $D_{MCMC}^\prime$ and $D_{DRL}^\prime$  are the new generated datasets  by applying the four different strategies on  $D_I$ again ( $|D_{I}| = |D_{RS}^\prime|= |D_{GA}^\prime| = |D_{MCMC}^\prime| = |D_{DRL}^\prime|$). Then, we use the following datasets as the \textbf{testing dataset} ($D_{RS}, D_{GA}, D_{MCMC}, D_{DRL}$), which are the same as those used in \S \ref{subsec:effect_clone_detection_approach}. Due to randomness of these strategies,  the new generated datasets (e.g., $D_{RS}^\prime$) will not be identical to the testing datasets (e.g., $D_{RS}$).}
We re-train the models of the ML-based detectors without altering any parameters. 

\input{origainl_adversal}


We also re-generate the $P$,$R$,$F_1$ values of each detector against the  transformation strategies in Tab.~\ref{tab:adversarial_recall_e}. For example,   
the row $D_I$ in Tab.~\ref{tab:adversarial_recall_e}  represents the accuracy of the retained models on the test dataset $D_I$. Compablack with  the  row $D_I$ in Tab.~\ref{tab:tokenlstmPer},  for \TokenLSTM, the adversarial learning improves $P$, $R$, $F_1$ from 0.991 to 0.995. For \Astnn, adversarial training slightly improves its accuracy, increasing  $F_1$ from 0.977 to 0.979.  
For \TBCCD, the new training results in a slight decrease in $P$  and $F1$ to 0.981 and 0.983, $R$ slight increase to 0.984. \rev{The results are reasonable, as the adversarial samples are quite different from the original samples. Besides, the accuracy in Tab.~\ref{tab:tokenlstmPer} is very high, we cannot expect much improvement via the adversarial training.}

\input{operatorCombine}

\rev{Regarding the effectiveness of different strategies in escaping from the detection, the dataset of clone pairs generated by the DRLSG strategy (i.e., $D_{DRL}$) achieves the worst detection accuracy among the last four rows in  Tab.~\ref{tab:adversarial_recall_e} --- this result clearly shows clone pairs generated by the DRLSG strategy are least detectable, and at least \rev{13\%} of its generated clone pairs cannot be detected by \Astnn and \TBCCD. On the other side, the RS strategy (i.e., $D_{RS}$) seems to be the least effective among the four strategies, generating the most detectable clone pairs and resulting in high detection accuracy. Last, the strategies GA  (i.e., $D_{GA}$) and MCMC (i.e., $D_{MCMC}$) show similar effectiveness, achieving the accuracy  somewhere in between $D_{DRL}$ and $D_{RS}$}.


\rev{Regarding the robustness of different ML-based detectors in capturing various generated clones, we find that the accuracy  of these detectors is generally satisfactory after adversarial training. As shown in Fig.~\ref{fig:adver}, \emph{in general, adversarial training improves the robustness of these ML-based detectors by increasing their  $P$,$R$,$F_1$  values}.} To be specific, the $F_1$ value of \TokenLSTM has improved by \textcolor{black}{0.236} on average among all the strategies. In the best case, it increase the $F_1$ by \textcolor{black}{0.439} on the $D_{DRL}$ dataset. For \Astnn, the average increase of $F_1$ is \textcolor{black}{0.266}, and it achieves a maximum increase of  \textcolor{black}{0.337} on the  $D_{DRL}$ dataset. The last detector, \TBCCD,  and it achieves an {increase } of only \textcolor{black}{0.085}  on average, maximum increase of 0.210  on the  $D_{GA}$ dataset.

Now, we answer the \textbf{RQ2}: after the adversarial training with samples from \ourtool, the $F_1$ values for the ML-based clone detectors \emph{significantly} increase, indicating that the adversarial training  has enhanced the robustness of the ML-based detectors. Meanwhile, the DRLSG strategy has exhibited the best effectiveness in generating undetectable clones as it  makes  \TokenLSTM, \Astnn, and \TBCCD achieve the lowest accuracy among the four strategies.

\input{traditionalMethod}

\subsection{RQ3: Effectiveness of    Transformation Operators}
\label{sec:eval:op}

In this section, we answer the \textbf{RQ3} about the effectiveness of these atomic transformation operators. 
\rev{Different from the previous experiments using all operators, we  apply and assess these operators group by group.} In particular,
we divide our operations into two groups~\cite{xu2017secure}: (1)  those \rev{highly-relevant} to \rev{semantic clones}   (Op1-ChRename, Op2-ChFor, Op3-ChWhile, Op4-ChDo, Op5-ChIfElseIF, Op6-ChIf, Op7-ChSwitch, Op8-ChRelation, Op9-ChUnary, Op10-ChIncrement,  Op12-ChDefine and Op14-ChExchange), and (2)  those of simple obfuscations (Op13-ChAddJunk, Op11-Constant, Op15-ChDelete).

To explore the impact of these operations on the clone detectors, \rev{we generate two new clone datasets:  the clone dataset ($D_{Sem}$) that  applies only the first group of operators and the dataset ($D_{Obf}$) that  applies only the second group. }
\rev{To eliminate possible side-effects of different transformation strategies, we do not use any above heuristic strategies. }
Inside, our method is straightforward in that it aggressively sets all bits in the bit vectors to be 1 for certain transformation operators, resulting in the complete set of code transformations at all qualified code locations.

The column 'Original'  in Tab.~\ref{tab:opconmbine} indicates that the model is trained on the $D_I$ dataset, and the column 'Adversarial'  indicates that the model is trained by adding adversarial samples ({$D_I$+ 25\%$D_{RS}^\prime$+ 25\%$D_{GA}^\prime$+25\%$D_{MCMC}^\prime$+25\%$D_{DRL}^\prime$}). The $D_{I}$ row in Tab.~\ref{tab:opconmbine} shows the accuracy on the original dataset, and the $D_{Sem}$, $D_{Obf}$ indicate the accuracy on the dataset generated by the two types of transformation operations. 

\input{traditionalToolsTime}
\rev{In general, the accuracy of the model is significantly affected by the two groups of transformation operators --- no matter on column  'Original'   or 'Adversarial',  the $F_1$ values of the second row ($D_{Sem}$) and third ($D_{Obf}$)    are always decreased, in contrast with the first row ($D_I$). Between the two different groups of transformation, obfuscation-like transformations (e.g., $D_{Obf}$) are  better in bypassing the detection of 
\TokenLSTM, \Astnn, \TBCCD, with average F1 values of \textcolor{black}{0.742} and \textcolor{black}{0.945} in 'Original' training and 'Adversarial' training, respectively. Besides, transformations for semantic clones  $D_{Sem}$  are also valid in escaping from the  ML-based detection, with the average F1 values of \textcolor{black}{0.877} and \textcolor{black}{0.962}. Hence, the experiments prove that the obfuscation transformation operators are in general more effective in affecting the accuracy of the ML-based clone detectors.}

Now, we answer the \textbf{RQ3}: from the experimental results, both groups of transformation operators are effective in generating clone pairs that can escape from the ML-based clone detectors. For these two groups, obfuscation-like transformation operators are more effective  and \rev{lower more detection accuracy of \TextLSTM and \Astnn.}

\subsection{RQ4: Traditional Detectors vs. \ourtool}
\label{sec:eval:traditional}

In this section, we answer the \textbf{RQ4}, which explores \xyx{whether or not traditional clone detectors can defend \ourtool}. While ML-based detectors have achieved
desirable performance, traditional detectors are still widely used in practice. 
\rev{According to the previous study \cite{Huihui2017},  traditional clone detector \Deckard achieves a poor recall (0.05)  on the original OJClone dataset and \SourcererCC obtains a reasonable recall (0.74) on the original OJClone dataset.}   

To explore to what extent the cloned code from \ourtool may escape from traditional detectors, we 
conduct a set of experiments with four open-source detectors as \SourcererCC~\cite{SajnaniSSRL16}, \Deckard~\cite{JiangMSG07}, \CCAligner~\cite{WangSWXR18}, 
\NICARD~\cite{Chanchal2008}. 
Tab.~\ref{tab:traditionalMethodRecallE} shows the experimental results under the four different
strategies. 
\xyx{Tab.~\ref{tab:traditionalTime}  lists only the  {detection} time of the four detectors, as traditional clone detectors do not require a pre-processing or training stage and could perform clone detection directly. \rev{Notably, column 'Processing Number' refers to the number of threads we run each detector in parallel to speed up the detection. In practice, we run 104 threads (the same as the number of folders in OJClone). If not using parallel detection, the traditional detectors would take up to days or even one week to finish the detection.}}

Traditional detectors have no embedded ML  models. Instead, they detect upon the recognition of suspicious clone patterns within the source code representation like token~\cite{WangSWXR18,SajnaniSSRL16}, text~\cite{Chanchal2008}, tree~\cite{JiangMSG07}, dependency graph~\cite{Zou2021} and so on.  
We treat them as black boxes, which are fed with the set of generated code clone pairs from \ourtool and output the "clones or not" decision. Finally, we calculate the overall \emph{Recall} values for each detector.

Overall, \ourtool makes the majority of the traditional detectors expose low \emph{recall} 
values, with an average value of 0.049 for $D_{DRL}$. \xyx{\emph{Besides, the DRLSG strategy is more effective than other strategies in beating the traditional detectors}, since more code changes bring more code syntactic differences.} This means that, under the 
DRLSG strategy, \ourtool can quickly generate many code clone pairs 
that can bypass these detectors with more than a 90\% success rate.
We examine the detectors whose \emph{recall} values are below 0.10. \SourcererCC is a 
token-based detector that has the lowest \emph{recall} (0.001) and can be easily defeated. 
\NICARD uses flexible, pretty-printing, and code normalization 
to accurately detect intentional clones. It has the second-lowest \emph{recall} (0.026). \CCAligner uses a combination of sliding windows and hashes to 
detect large-gap type clones. Its \emph{recall} is 0.07. \Deckard is a tree-based clone 
detection tool with a 0.099 \emph{recall} value.

%

Now, we answer the \textbf{RQ4}:  
\xyx{the traditional clone detectors have no resistance to the clone pairs generated by \ourtool, especially to the clone pairs in $D_{DRL}$.  For the tree-based detector  (i.e., \Deckard) that is believed to be more robust than token-based ones, its detection recall is also less than 10\% on the datasets \ourtool generates.}

\section{Discussion}

\subsection{Threats to Validity}
{Threats to internal validity come from the parameter setup for the traditional and the ML-based detectors used in this paper. To address this issue, for the traditional detectors, we use their default settings. Since traditional detectors usually have few parameters to tune, their results are quite stable. For the ML-based detectors, we use the same parameters as those reported in their papers \cite{YuLCLXW19,Zhang2019}. We also fine-tune the unreported parameters and finally make the trained models reproduce similar results as those reported in the corresponding papers. Through this rigorous process, we believe that we have conducted fair comparisons between our approach and all the baseline clone detectors in this study. 
}

{Threats to external validity mainly come from two aspects. First, we just use the OJClone C language dataset to evaluate the detectors, 
which is a widely-used open-source benchmark for large-scale semantic clone detection. Currently, we have only done evaluations on C/C++ programs, however, like other high-level languages (e.g., Java) that share similar language features (e.g., object-orientation), we believe the evaluation findings could also be generalized to clone detection for other high-level languages. Second, we evaluate only three ML-based detectors in this paper due to the tool availability issue. In the future, we would like to support more ML-based detectors (e.g., \cite{zhang2019find,zhao2018deepsim} when they are publicly available) or even code plagiarism tools such as \textsc{Moss}~\cite{schleimer2003winnowing}. }


\subsection{Impact of Transformation Operators}
\label{sub:theAttrAffect}

 \input{mutationInfluenceAttributes}

As shown in Tab.~\ref{tab:MutationInfluence}, we have analyzed the impact of 15 proposed transformation operators on the four commonly-used representations of source code. The impact could be categorized into three levels.  
Firstly, severe impact (denoted by \priority{100}) means that the operator breaks the original structure of the representation, while minor impact (denoted by \priority{50})  means that it only changes the node properties of the representation (e.g., changing one node in \emph{AST} or \emph{CFG}). Last, no impact (denoted by \priority{0}) means that neither the properties nor the structure of the representation is modified.  

{Token-based detectors are susceptible to all the changes bought by the transformation operators, except Op1 (identifier renaming) and Op14 (statement exchanging).  
Hence, token-based detectors can only detect Type I and Type II clones, and our operators can easily evade the token-based detectors (e.g., \SourcererCC~\cite{SajnaniSSRL16}, \CCAligner~\cite{WangSWXR18}, \CCFinderX~\cite{Toshihiro2002}). AST-based detectors (e.g., \Deckard~\cite{JiangMSG07}) perform better than the token-based ones, but the AST structure is not resilient to control- or data-flow changes made by operators such as  Op2 to Op14. Therefore, AST-based detectors are good at detecting Type I and II clones and these detectors could be evaded by our operators. 
In contrast, CFG- or PDG-based detectors (e.g., \CCGraph~\cite{Zou2021}) are more resilient to the  control- or data-flow changes. Especially for CFG-based detectors, they are more resilient to the transformation performed by Op8 to Op11 (e.g., changing \codeff{while} to \codeff{for-loop}). 
However, the Op4, Op7, Op12, and Op13 would have a severe impact on the four representations, as the transformations brought by these operators belong to Type-IV clones (i.e., semantic clones). 
}

\subsection{Lightweight Code Transformation or Obfuscators?}

In this study, we use transformation operators and strategies and implement our framework \ourtool to perform the lightweight \xyx{semantic-preserving} transformation. Software obfuscators~\cite{pawlowski2016probfuscation,liu2017stochastic,wang2016translingual,picheta2020code,laszlo2009obfuscating} can also enable equivalence transformations of source code.  
The reasons why we do not employ existing obfuscators in this work are threefold:  
1) defeating clone detection is essentially a trade-off problem that strikes the balance between costs in code transformation and benefits from evasion. Obfuscators are generally not free of charge and \xyx{comparatively time-consuming}. Hence, we offer a lightweight yet effective code transformation approach. 
2)  Obfuscation is often utilized to protect the IP (Intellectual Property). Therefore, the code after obfuscation often has poor readability and maintainability. \xyx{For example, for the code \codeff{int i = 1;}, after the encoding by the obfuscator~\cite{picheta2020code}, it will become:}
 
\vspace{2mm}
\begin{lstlisting}  [ 
		language=c++,
		backgroundcolor=\color{white},
		extendedchars=true,
	    basicstyle=\fontsize{8}{8}\linespread{0.8}\selectfont\ttfamily,%\footnotesize\ttfamily,
		%basicstyle=\footnotesize,
		numbers=none,
		numberstyle=\tiny\color{gray},
		numberstyle=\color{gray},
		keywordstyle=\color{blue}, 
		xleftmargin=.25in,
		breaklines=true,
		captionpos=b,
		keepspaces=true,
		stepnumber=1,
		aboveskip=1mm,
		belowskip=0.5mm]
int o_8ffc9af5e5913588bc0b7705602caf02= (0x0000000000000002 + 0x0000000000000201 + 0x0000000000000801 - 0x0000000000000A03);
\end{lstlisting} 
\vspace{2mm}

3) More importantly, small rather than complicated transformations are much easier for algorithm developers to analyze and debug robustness issues. Hence, 
simple yet effective transformations are always favorable to algorithm developers.


%% file: clonePairsAndGenTime.tex
%

\begin{table}[t]
	\footnotesize
	\caption{The code generation time}
	\label{tab:dataPairsAndGenTiem}
	\vspace{-2mm}
	\centering
	\setlength{\tabcolsep}{5.5mm}{
		\begin{tabular}{ccccc}
			\toprule
			Time  & RS   & GA   & MCMC  & DRLSG \\ \midrule
			T(h) & 0.88 & 4.33 & 20.70 & 168 \\ 
			\bottomrule
		\end{tabular}
	}
	\vspace{-2mm}
	\centering
\end{table}

%% file: tokenLSTMPerformance.tex
\begin{table}[t]
	\footnotesize
	\caption{Detection accuracy of the ML-based detectors on OJClone}
	\label{tab:tokenlstmPer}
	\vspace{-2mm}
\renewcommand\arraystretch{1.3} 
	\setlength{\tabcolsep}{6.45mm}{
	\begin{tabular}{lccc}
		\toprule 
		{\color[HTML]{000000} Tools}      & {\color[HTML]{000000} P}     & {\color[HTML]{000000} R}     & {\color[HTML]{000000} F1}    \\ \midrule
		{\color[HTML]{000000} \TokenLSTM}  & {\color[HTML]{000000} \textbf{0.991}} & {\color[HTML]{000000} \textbf{0.991}} & {\color[HTML]{000000} \textbf{0.991}} \\
		{\color[HTML]{000000} \Astnn}      & {\color[HTML]{000000} 0.996} & {\color[HTML]{000000} 0.960} & {\color[HTML]{000000} 0.977} \\
		{\color[HTML]{000000} \TBCCD}      & {\color[HTML]{000000} 0.987} & {\color[HTML]{000000} 0.994} & {\color[HTML]{000000} 0.990} \\
		\bottomrule
	\end{tabular}
}
\vspace{-2mm}
\end{table}

%% file: dlTrainTestTime.tex
\begin{table}[t]
\footnotesize
\caption{Time consumption of the ML-based clone detectors}
 \vspace{-2mm}
 \renewcommand\arraystretch{1.3} 
\label{tab:dlTrainTest}
\centering
\setlength{\tabcolsep}{2.158mm}{
\begin{tabular}{p{2.8cm}p{2.1cm}p{2.6cm}}
\toprule
Tools & \begin{tabular}[c]{@{}c@{}}Pre-processing(s) \end{tabular} & \begin{tabular}[c]{@{}c@{}}Training \& Testing(s)\end{tabular} \\ 
\midrule
\TokenLSTM &  656 & 21963   \\ 
\Astnn     &  2412 & 7419   \\
\TBCCD     &  4991 & 187467 \\
\bottomrule
\end{tabular}
}
\vspace{-2mm}
\centering
\end{table}

%% file: dlAllTestResult.tex
\begin{table}[t]
\footnotesize
\caption{Results of the ML-based detectors in detecting clone pairs generated by \ourtool}
\vspace{-2mm}
\renewcommand\arraystretch{1.3}  
\label{tab:oringinal_recall_e}
\centering
\setlength{\tabcolsep}{1.02mm}{
\begin{tabular}{lccccccccc}
\toprule
{\color[HTML]{000000} } &
\multicolumn{3}{c}{{\color[HTML]{000000} \TokenLSTM}} &
\multicolumn{3}{c}{{\color[HTML]{000000} \Astnn}} &
\multicolumn{3}{c}{{\color[HTML]{000000} \TBCCD}} \\ \midrule
TestData &
\multicolumn{1}{c}{{\color[HTML]{000000} P}} &
\multicolumn{1}{c}{{\color[HTML]{000000} R}} &
\multicolumn{1}{c}{{\color[HTML]{000000} F1}} &
\multicolumn{1}{c}{{\color[HTML]{000000} P}} &
\multicolumn{1}{c}{{\color[HTML]{000000} R}} &
\multicolumn{1}{c}{{\color[HTML]{000000} F1}} &
\multicolumn{1}{c}{{\color[HTML]{000000} P}} &
\multicolumn{1}{c}{{\color[HTML]{000000} R}} &
\multicolumn{1}{c}{{\color[HTML]{000000} F1}} \\ \hline
{\color[HTML]{000000}  $D_{RS}$} &
{\color[HTML]{000000} 0.891} &
{\color[HTML]{000000} 0.882} &
{\color[HTML]{000000} 0.882} &
{\color[HTML]{000000} 0.803} &
{\color[HTML]{000000} 0.721} &
{\color[HTML]{000000} 0.701} &

{\color[HTML]{000000} 0.929} &
{\color[HTML]{000000} 0.855} &
{\color[HTML]{000000} 0.891} \\
{\color[HTML]{000000}  $D_{GA}$} &
{\color[HTML]{000000} 0.713} &
{\color[HTML]{000000} 0.679} &
{\color[HTML]{000000} 0.666} &
{\color[HTML]{000000} 0.770} &
{\color[HTML]{000000} 0.641} &
{\color[HTML]{000000} 0.592} &

{\color[HTML]{000000} 0.822} &
{\color[HTML]{000000} 0.671} &
{\color[HTML]{000000} \textbf{0.739}} \\
{\color[HTML]{000000}  $D_{MCMC}$} &
{\color[HTML]{000000} 0.888} &
{\color[HTML]{000000} 0.878} &
{\color[HTML]{000000} 0.877} &
{\color[HTML]{000000} 0.800} &
{\color[HTML]{000000} 0.703} &
{\color[HTML]{000000} 0.676} &

{\color[HTML]{000000} 0.873} &
{\color[HTML]{000000} 0.947} &
{\color[HTML]{000000} 0.908} \\
{\color[HTML]{000000} $D_{DRL}$} &
{\color[HTML]{000000} 0.607} &
{\color[HTML]{000000} 0.558} &
{\color[HTML]{000000} \textbf{0.502}} &
{\color[HTML]{000000} 0.739} &
{\color[HTML]{000000} 0.599} &
{\color[HTML]{000000} \textbf{0.530}} &

{\color[HTML]{000000} 0.854} &
{\color[HTML]{000000} 0.871} &
{\color[HTML]{000000} 0.863} \\ \bottomrule
\end{tabular}
}
 \vspace{-2mm}
\centering
\end{table}

%% file: dlMutatedAddedTestResult.tex

\begin{table}[t]
	\footnotesize
		\caption{Results of the ML-based detectors \textbf{with adversarial training} in detecting clone pairs generated by \ourtool}
	 \vspace{-2mm}
	\label{tab:adversarial_recall_e}
	\centering
	\renewcommand\arraystretch{1.2}  
	\setlength{\tabcolsep}{1mm}{
	\begin{tabular}{lccccccccc}
	\toprule
	{\color[HTML]{000000} }  &
	\multicolumn{3}{c}{{\color[HTML]{000000} \TokenLSTM}} &
	\multicolumn{3}{c}{{\color[HTML]{000000} \Astnn}} &
	\multicolumn{3}{c}{{\color[HTML]{000000} \TBCCD}} \\ \hline
	 TestData &  P & R & F1 & P& R & F1& P& R & F1 \\ \midrule
	{\color[HTML]{000000} $D_I$} &
	{\color[HTML]{000000} 0.995} &
	{\color[HTML]{000000} 0.995} &
	{\color[HTML]{000000} 0.995} &
	{\color[HTML]{000000} 0.994} &
	{\color[HTML]{000000} 0.965} &
	{\color[HTML]{000000} 0.979} &
	
	{\color[HTML]{000000} 0.981} &
	{\color[HTML]{000000} 0.984} &
	{\color[HTML]{000000} 0.983} \\
	{\color[HTML]{000000} $D_{RS}$} &
	{\color[HTML]{000000} 0.989} &
	{\color[HTML]{000000} 0.989} &
	{\color[HTML]{000000} 0.989} &
	
	{\color[HTML]{000000} 0.922} &
	{\color[HTML]{000000} 0.920} &
	{\color[HTML]{000000} 0.920} &
	
	{\color[HTML]{000000} 0.948} &
	{\color[HTML]{000000} 0.974} &
	{\color[HTML]{000000} 0.961} \\
	{\color[HTML]{000000} $D_{GA}$} &
	{\color[HTML]{000000} 0.954} &
	{\color[HTML]{000000} 0.952} &
	{\color[HTML]{000000} 0.952} &
	
	{\color[HTML]{000000} 0.891} &
	{\color[HTML]{000000} 0.880} &
	{\color[HTML]{000000} 0.879} &

	{\color[HTML]{000000} 0.928} &
	{\color[HTML]{000000} 0.971} &
	{\color[HTML]{000000} 0.949} \\
	{\color[HTML]{000000} $D_{MCMC}$} &
	{\color[HTML]{000000} 0.989} &
	{\color[HTML]{000000} 0.989} &
	{\color[HTML]{000000} 0.989} &
	
	{\color[HTML]{000000} 0.901} &
	{\color[HTML]{000000} 0.897} &
	{\color[HTML]{000000} 0.897} &

	{\color[HTML]{000000} 0.921} &
	{\color[HTML]{000000} 0.976} &
	{\color[HTML]{000000} 0.961} \\
	{\color[HTML]{000000} $D_{DRL}$} &
	
	{\color[HTML]{000000} 0.943} &
	{\color[HTML]{000000} 0.941} &
	{\color[HTML]{000000} \textbf{0.941}} &
	
	{\color[HTML]{000000} 0.881} &
	{\color[HTML]{000000} 0.868} &
	{\color[HTML]{000000} \textbf{0.867}} &

	{\color[HTML]{000000} 0.903} &
	{\color[HTML]{000000} 0.842} &
	{\color[HTML]{000000} \textbf{0.872}} \\ 
	\bottomrule
	\end{tabular}
}
\vspace{-4mm}
\end{table}

%% file: origainl_adversal.tex
\begin{figure}[t]
	\centering
	\includegraphics[width=\linewidth]{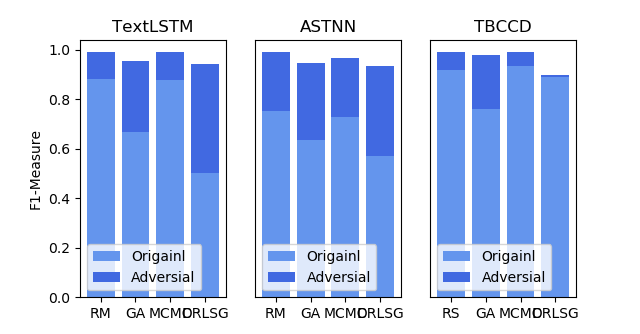}
		\vspace{-4mm}
	\caption{F-measure before and after adversarial training} 
	 	\vspace{-4mm}
	\label{fig:adver}
\end{figure}

%% file: operatorCombine.tex
\begin{table*}[!thbp]
	\caption{The effectiveness of the transformation operator.}
	\label{tab:opconmbine}
	\vspace{-2mm}
	\renewcommand\arraystretch{1.2}  
	\footnotesize
	\setlength{\tabcolsep}{1.5mm}{
	\begin{tabular}{l|cccccc|cccccc|cccccc}
		\hline
		\multicolumn{1}{c|}{} &
		\multicolumn{6}{c|}{\TokenLSTM}               & \multicolumn{6}{c|}{\Astnn}                    & 
		\multicolumn{6}{c}{\TBCCD}                    \\ \hline
		\multicolumn{1}{c|}{} &
		\multicolumn{3}{c}{Original} &
		\multicolumn{3}{c|}{Adversial} &
		\multicolumn{3}{c}{Original} &
		\multicolumn{3}{c|}{Adversial} &
		\multicolumn{3}{c}{Original} &
		\multicolumn{3}{c}{Adversial} \\ \hline
		TestData & P     & R     & F1    & P     & R     & F1    & P     & R     & F1    & P     & R     & F1    & P     & R     & F1    & P     & R     & F1    \\ \hline
		$D_I$& 
		0.991 & 0.991 & 0.991 & 0.995 & 0.995 & 0.995 &
		0.996 & 0.960 & 0.977 & 0.994 & 0.965 & 0.979 &
		0.987 & 0.994 & 0.990 & 0.983 & 0.983 & 0.983 \\
		
		$D_{Sem}$    & 
		0.899 & 0.891 & 0.890 & 0.995 & 0.995 & 0.995 &
		0.856 & 0.817 & 0.811 & 0.927 & 0.927 & 0.927 & 
		0.941 & 0.918 & 0.929 & 0.957 & 0.971 & \textbf{0.964} \\
		
		$D_{Obf}$      &
		0.730 & 0.701 & \textbf{0.691} & 0.969 & 0.969 & \textbf{0.969} &
		0.787 & 0.700 & \textbf{0.675} & 0.905 & 0.901 &  \textbf{0.901} &
		0.864 & 0.856 & \textbf{0.860} & 0.940 & 0.989 & \textbf{0.964} \\ \hline
	\end{tabular}}
	\vspace{-2mm}
\end{table*}

%% file: traditionalMethod.tex
\begin{table}[t]   
\footnotesize
\caption{The \emph{Recall} values of traditional clone detectors}     
\label{tab:traditionalMethodRecallE} 
 \vspace{-2mm}
\centering    
\renewcommand\arraystretch{1.2} 
\setlength{\tabcolsep}{1.95mm}{
\begin{tabular}{lcccc}
\toprule
TestData & \SourcererCC & \NICARD & \CCAligner & \Deckard \\
\midrule
$D_{RS}$ & 0.006 & 0.027 & 0.100 & 0.110 \\
$D_{GA}$ & 0.004 & \textbf{0.026} & 0.076 & 0.101 \\
$D_{MCMC}$ & 0.008 & 0.046 & 0.083 & 0.100\\
$D_{DRL}$ & \textbf{0.001} & 0.027 & \textbf{0.070} & \textbf{0.099} \\
\bottomrule

\end{tabular}
}
 \vspace{-1mm}
\centering 
\end{table}


%% file: traditionalToolsTime.tex
\begin{table}[t]
\footnotesize
\caption{Time consumption of the traditional clone detectors}
 \vspace{-2mm}
\label{tab:traditionalTime}
\centering
\renewcommand\arraystretch{1.2} 
\setlength{\tabcolsep}{2.26mm}{
\begin{tabular}{p{2.1cm}p{2.8cm}p{2.6cm}}
\toprule
Detector & \begin{tabular}[c]{@{}c@{}}Processing Number \end{tabular} & \begin{tabular}[c]{@{}c@{}} Times(m)\end{tabular} \\ 
\midrule

\SourcererCC & 104&  118 \\
\NICARD  & 104&   0.16\\
\CCAligner & 104 &   12  \\
\Deckard & 104 & 221 \\
\bottomrule
\end{tabular}
}
 \vspace{-4mm}
\centering
\end{table}

%% file: mutationInfluenceAttributes.tex
\begin{table}[!h]
\footnotesize
\begin{center}
\caption{The transformation operators and their impact on the four commonly-used representations of the code} 
\renewcommand\arraystretch{1.2} 
 \vspace{-1mm}
\label{tab:MutationInfluence}

\setlength{\tabcolsep}{3.6mm}{
\begin{tabular}{p{2.7cm}cccc}
  \toprule
   Operator  &AST & CFG & PDG & Token \\
  \midrule
  Op1-ChRename 
  &\priority{50} 
  &\priority{0} 
  & \priority{50}
  & \priority{0}\\ 
  Op2-ChFor
  &\priority{100} 
  &\priority{0} 
  & \priority{0} 
  & \priority{100}\\ 
  Op3-ChWhile 
  &\priority{50} 
  &\priority{0} 
  & \priority{0}
  & \priority{100}\\  
  Op4-ChDo
  &\priority{100} 
  &\priority{100} 
  & \priority{100} 
  & \priority{100}\\  
  Op5-ChIfElseIF
  &\priority{50}
  &\priority{0} 
  &\priority{0} 
  & \priority{100} \\ 
  Op6-ChIf 
  &\priority{50}
  &\priority{0}
  &\priority{0} 
  & \priority{100}\\ 
  Op7-ChSwitch
  &\priority{100} 
  &\priority{100} 
  & \priority{100} 
  &\priority{100}\\  
  Op8-ChRelation 
  &\priority{100} 
  &\priority{0} 
  &\priority{50} 
  & \priority{100}\\  
  Op9-ChUnary 
  &\priority{100}
  &\priority{0} 
  &  \priority{50}
  & \priority{100}\\  
  Op10-ChIncrement 
  &\priority{100} 
  &\priority{0}
  &\priority{50} 
  &\priority{100}\\  
  Op11-ChConstant  
  &\priority{100} 
  &\priority{0}
  & \priority{50}
  & \priority{100}\\  
  Op12-ChDefine
  &\priority{100} 
  &\priority{100} 
  &  \priority{100}
  & \priority{100}\\  
  Op13-ChAddJunk 
  &\priority{100} 
  &\priority{100} 
  &\priority{100}
  & \priority{100}\\  
  Op14-ChExchange  
  &\priority{0}
  &\priority{100} 
  & \priority{0}
  & \priority{50}\\  
  Op15-ChDeleteC 
  &\priority{100} 
  &\priority{50} 
  & \priority{50} 
  & \priority{100}\\ 
  \bottomrule
\end{tabular}
}
\end{center}
{\raggedright $^*$Note that we use the symbol \lq\lq{}\priority{100}\rq\rq to denote severe effects,  \lq\lq{}\priority{50}\rq\rq  to denote only minor effects, and   \lq\lq{}\priority{0}\rq\rq to denote no effects.  \par}
 \vspace{-2mm}
\end{table}

%% file: related.tex
\section{Related Work}
\label{sec:related}

\subsection{Code Clone Detection}
\label{sec:mtv:clonetype}
Generally, existing clone detectors can be classified as textual-based, token-based, structural-based code cloning detection approaches.
The textual-based methods~\cite{seunghak2005,Chanchal2008,kim2017vuddy} represent code fragments in the form of strings. If the text contents of the two code fragments are similar, they are considered clones.
The approaches described in~\cite{Zhenmin2006,WangSWXR18,Liuqing2017,SajnaniSSRL16,Toshihiro2002}  represent source code as a series of token sequences and use different similarity detection algorithms on the token sequences to detect code clones. 
The approaches described in~\cite{JiangMSG07,Zhang2019,baxter1998clone,wei2017supervised} detect code similarity by extracting the syntax of the code to obtain the semantic features of the code. 
Recently, some approaches~\cite{Liu2006,zhao2018deepsim,wang2017ccsharp} perform code clone detection by extracting semantic code features, for example using hybrid features, such as CFGs and ASTs of a program ~\cite{FangLS0S20,sheneamer2016semantic,ragkhitwetsagul2018comparison}, or through learning-based approaches
~\cite{FangLS0S20,Zhang2019,YuLCLXW19,Liuqing2017,wei2018positive,zhao2018deepsim,wei2017supervised,sheneamer2016semantic,white2016deep,hua2020fcca}.
Regarding the assessment of clone detectors, \rev{BigCloneBench~\cite{2015Evaluating} provide a clone detection benchmark to evaluate the clone detectors, and existing studies~\cite{roy2009comparison,1134103,5279980,1342759,BellonKAKM07,6976098} focus on evaluating traditional detectors in certain aspects, and there is still a lack of studies that systematically challenge and assess the robustness of the recent ML-based clone detectors}.
 
\subsection{Code Mutation}
Code mutations are often used in code testing. Mutation testing, which generates a large number of mutants that are automatically embedded in the code to exercise a target program to detect its  bugs~\cite{Jia2011,Yao2014,ZhangZHHJZ19,McMinn2019}. 
Mutation testing can also be used to test the effectiveness of code clone detectors. 
Roy et al. \cite{Chanchal2008} identified and standardized potential clones, and then used dynamic clustering to perform simple text-line comparisons of potential clones.
Roy et al.~\cite{RoyC09} proposed a mutation insertion method to test code clones. The idea is to reinsert an artificial piece of code into a piece of source code so that different types of code clone pairs can be artificially forged and then tested against the target clone detectors. Their proposed tool is a random transformation of the code, which may change the semantics of the original code. 
Svajlenko et al.~\cite{Svajlenko2019} presented a benchmark framework based on mutation analysis, which evaluates the recall rate of clone detection tools for different types of clones and the editing of specific types of clones does not require human intervention. Unlike approaches~\cite{Svajlenko2019,RoyC09} that can transform code but cannot guarantee semantic preservation, our approach always generates semantic equivalent clones for robustness validation. \rev{Recently, Zhang et al.~\cite{zhang2020generating} only used one transformation operator (renaming variable), which is included in this work, to prove that some source code processing methods (e.g., \Astnn and Token-based LSTM models) are not sound in the code classification problem. In this paper, our approach proves that the ML-based clone detectors are not sound enough in detecting    code clones after simple yet effective equivalent transformations.}

\subsection{Code Obfuscation}
\rev{Equivalence transformations are also commonly used in code obfuscation. Program obfuscation is a set of semantic-preserving program mutation techniques. It is mainly used to hide the intent of a program or to protect the intellectual property of software before its release. Liu et al.~\cite{liu2017stochastic} proposed a  language-model-based obfuscation framework. It makes code refactoring tools like JSNice~\cite{raychev2015predicting} more difficult to refactor a program. Breanna et al.~\cite{devore2020mossad} presented \textsc{Mossad}, a method for making code plagiarism tools by inserting junk code, which effectively defeats theft detectors such as \textsc{Moss}~\cite{schleimer2003winnowing}. 
{Schulze et al~\cite{schulze2013robustness} proposed to apply code obfuscations to evaluate the robustness of some traditional clone detectors. They applied a few code obfuscations semi-automatically to source code and did not consider strategies to guide code mutation.} 
The goal of our work is to conduct a simple yet effective transformation to generate semantic clones to evade both learning-based and traditional clone detectors. Rather than applying a heavy-weight transformation (encoding~\cite{picheta2020code}, CFG flatten~\cite{laszlo2009obfuscating} or other compiler optimizations~\cite{ren2021unleashing}), our lightweight approach makes it easier for developers to quickly discover and locate robustness issues in a clone detector. }

%% file: conclusion.tex
\section{Conclusion}
\label{sec:conclusion}
This paper presents \ourtool, a lightweight yet effective code transformation framework that can assess the robustness of ML-based clone detectors by automatically generating clone pairs. 
several state-of-the-art ML-based and traditional clone detectors. The experimental results show that our  lightweight transformations are effective in evaluating the robustness of clone detectors and can significantly reduce the performance of three recent ML-based detectors, i.e.,  \Astnn, \TBCCD, \TokenLSTM. 
Our study reveals the robustness implications of the machine learning-based clone detectors, which calls for more robust and effective methods for data collection and model training. One possible solution is to design a hybrid source code representation to improve the capability of existing ML-based detectors.
Our source code and experimental data are publicly available at {https://github.com/CloneGen/CLONEGEN}.